\documentclass[aps,prl,twocolumn]{revtex4-1}
\usepackage{graphicx}
\usepackage{dcolumn}
\usepackage{bm}
\usepackage{multirow}
\usepackage{ulem}
\usepackage{color}
\usepackage{booktabs}
\usepackage{amsmath}
\usepackage{hyperref}
\usepackage{cleveref}
\usepackage{textcomp}
\usepackage{makecell}
\usepackage{CJKutf8} 
\begin{document}
\title{Single-Atomic-Ensemble Dual-Wavelength Optical Standard}

\author{$^1$Jie Miao}
\author{$^1$Jingming Chen}
\author{$^2$Deshui Yu}
\author{$^1$Qiaohui Yang}
\author{$^1$Duo Pan}
\email{panduo@pku.edu.cn}
\author{$^{1,3}$Jingbiao Chen}

\affiliation{$^1$State Key Laboratory of Advanced Optical Communication Systems and Networks, Department of Electronics, Peking University, Beijing 100871, China}
\affiliation{$^2$National Time Service Center, Chinese Academy of Sciences, Xi'an 710600, China}
\affiliation{$^3$Hefei National Laboratory, Heifei 230088, China}

\begin{abstract}
We demonstrate a dual-wavelength optical frequency standard based on the dual-optical-transition modulation transfer spectroscopy (DOT-MTS) between different quantum transitions of the rubidium D1 (795 nm) and D2 (780 nm) lines. In a single rubidium atomic ensemble, modulation frequency sidebands from the 780 nm pump beam are simultaneously transferred to both the 780 nm and 795 nm probe lasers.  The DOT-MTS enables the simultaneous stabilization of 780 nm and 795 nm lasers on a single vapor cell. Both lasers exhibit a frequency instability in the low 10$^{-14}$ range at 1 s of averaging, as estimated from the residual error signal. A theoretical model is developed based on the V-type atomic level structure to illustrate the dual-wavelength spectroscopy. This approach can be extended to develop a multi-wavelength optical frequency standard within a single atomic ensemble, broadening its applicability in fields such as precision metrology, wavelength standards, optical networks, and beyond.
\end{abstract}
\date{\today}

\maketitle
Optical frequency standards referenced to thermal atoms, employing sub-Doppler spectroscopy techniques, including saturation absorption spectroscopy (SAS)\cite{SAS1,SAS2,SAS3}, polarization spectroscopy\cite{PS1,PS2,PS3}, two-photon spectroscopy\cite{TP1,TP2,TP3,TP4}, dual-frequency sub-Doppler spectroscopy\cite{DFSDS1,DFSDS2,DFSDS3}, frequency modulation spectroscopy\cite{FM1,FM2} and modulation transfer spectroscopy (MTS)\cite{MTS2,MTS1,MTS3,MTS4,MTS5}, have facilitated advancements in precision measurement \cite{pm1,pm2,pm3}, optical clocks \cite{pm1,ac2,ac3,ac4}, cold atom physics\cite{cd1,cd2}, and navigational systems\cite{nav1,nav2}. Benefiting from enhanced sensitivity and noise rejection, which yield a high SNR and superior frequency stability, MTS and frequency modulation spectroscopy are essential for the stabilization of compact laser systems\cite{com1}. They have been effectively applied to iodine\cite{I1}, acetylene\cite{ace1}, and alkali atoms\cite{alkali1,alkali2,alkali3}, achieving the best short-term instability in thermal atomic ensembles to date\cite{Best1,ac3,I1}. Utilizing the quantum transitions of iodine and acetylene molecules facilitates optical frequency standards at diverse wavelength regions, crucial for the realization of meter, precision metrology\cite{metro}, and optical telecommunication\cite{tele1,tele2}. In various application scenarios, such as constructing optical frequency networks\cite{network1,network2} and cold atom physics, the establishment of multi-wavelength frequency standards necessitates multiple independent quantum ensembles, increasing system complexity. Nevertheless, the potential for realizing multiple wavelength standards within a single quantum ensemble remains uninvestigated. Here, we implement modulation transfer spectroscopy between different quantum transitions to establish the first dual-wavelength optical frequency standard based on a single quantum ensemble.

Research on modulation transfer spectroscopy (MTS) has primarily focused on single quantum transitions\cite{Theory1,Theory2,Theory3,Theory4}, where both pump and probe lights originate from the same laser beam and modulation is transferred through the same transition, thereby limiting its application to single-quantum transition laser stabilization. Modulation transfer between different quantum transitions was preliminarily realized using Ne transitions\cite{DOTMTS1}, with the initial theoretical analysis in 1986 \cite{DOTMTS2}. After years of stagnation, in 2015, progress was made by transferring modulation from a prelocked 780 nm laser in a separate atomic vapor cell to a 1529 nm laser, resulting in the locking of a second atomic vapor cell at 1529 nm with a frequency stability of $3.3 \times 10^{-12} \sqrt{\tau/s}$, derived from the error signal \cite{DOTMTS3}. In 2022, following the work made in 2015, a four-level modulation transfer scheme involving 770 nm, 1529 nm, and 780 nm was locked with three independent atomic vapor cells independently for 780nm, 1529nm and 770nm\cite{DOTMTS4}. In this paper, we demonstrate dual-optical-transition modulation transfer spectroscopy (DOT-MTS) using solely a single atomic cell, where the modulation frequency sidebands of the pump beam, resonant with the $^{87}\text{Rb}$ D2 line (780 nm), are simultaneously transferred to both the 780 nm (resonant with the $^{87}\text{Rb}$ D2 line) and 795 nm (resonant with the $^{87}\text{Rb}$ D1 line) probe beams. 780 nm and 795 nm lasers are simultaneously stabilized on a thermal atomic ensemble through the modulation transfer. The resultant frequency stabilities reach the low 10$^{-14}$ level at 1 s of averaging, as estimated from the residual error signal. The approach enables the realization of a dual-wavelength or even a multi-wavelength frequency standard. And a theoretical model based on the V-type atomic level structure is developed to illustrate the dual-wavelength spectroscopy. To our knowledge, this work represents the first demonstration of dual-wavelength laser stabilization in a single atomic gas cell, enabling the realization of multiple quantum standards within the same system to enhance coherence, necessitating further investigation into the underlying fundamental physics and complex correlations.

\begin{figure*}[t]
	\includegraphics[width=17.5cm,height=11cm]{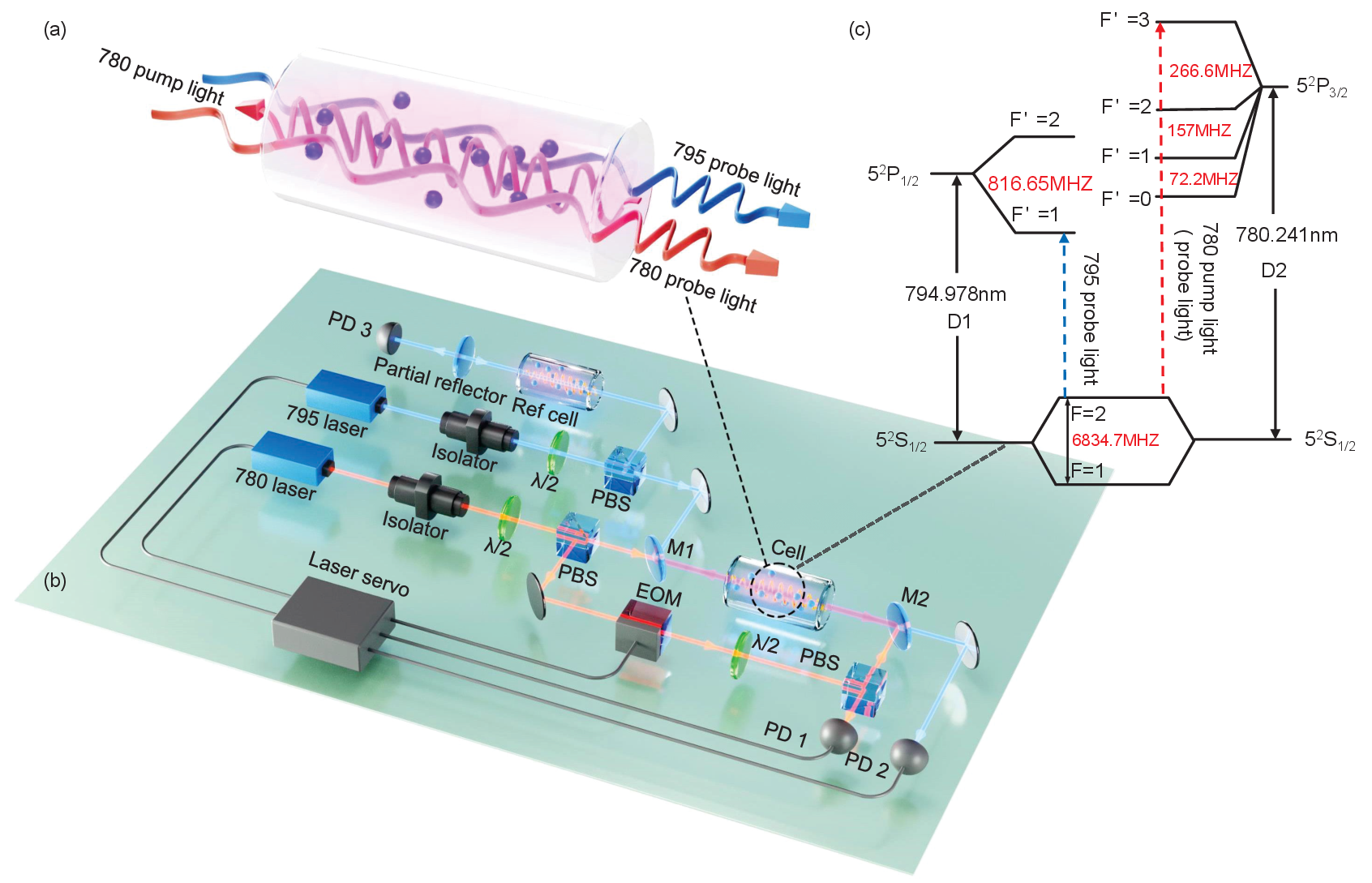}
	
	\caption{The 780-795 nm DOT-MTS. (a) The modulation transfer process in 780-795 nm DOT-MTS and 780 nm MTS: the modulation of 780 nm pump light is transferred to the 780 nm and 795 nm probe light in a $^{87}$Rb cell. (b) The optical setup of the 780-795 nm DOT-MTS and 780 nm MTS includes a 780 nm ECDL (external-cavity diode laser), 795 nm ECDL, isolator, $\lambda$/2 (half-wave plate), PBS (polarization beam splitter), dichroic mirror M1 (780 nm transmission, 795 nm reflection), dichroic mirror M2 (780 nm reflection, 795 nm transmission), $^{87}$Rb cell, EOM (electro-optic modulator), PD1 (photodiode detector 1), PD2, and laser servo. (c) The energy levels of 780 nm MTS and 780-795 nm DOT-MTS: the blue line corresponds to the 795 nm probe light, and the red line corresponds to the 780 nm pump light and probe light.}
	\label{setup}
\end{figure*}

\begin{figure*}[t]
	\includegraphics[width=17cm,height=4.54cm]{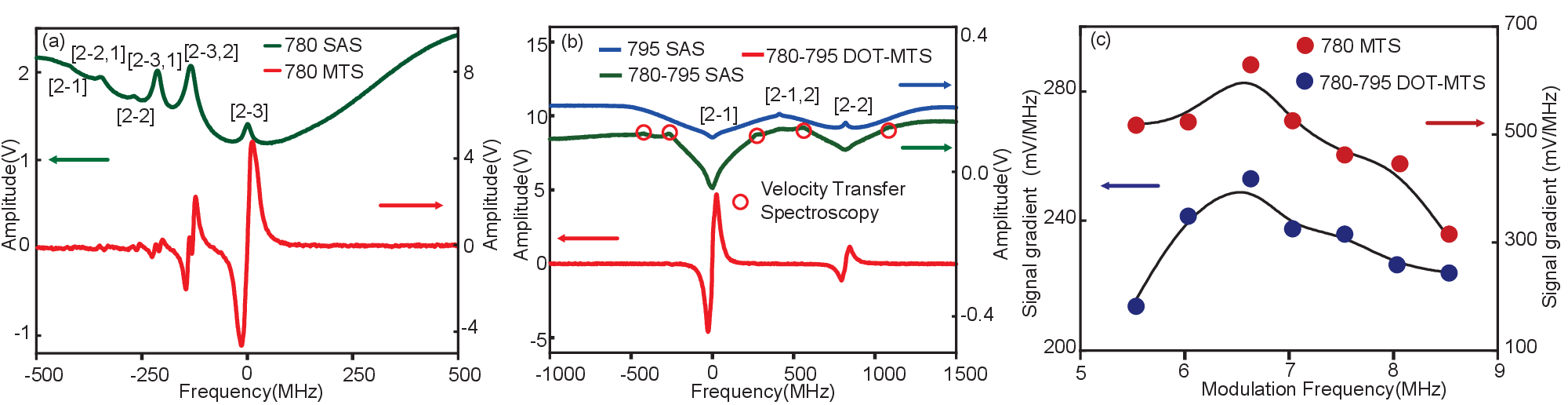}
	\caption{(a) 780 nm SAS and MTS: The green line represents the 780 nm SAS, showcasing three resonance peaks and three crossover peaks. The red line depicts the 780 nm MTS, with the 5$^{2}S_{1/2}$ F=2 $\rightarrow$ 5$^{2}P_{3/2}$ F=3 transition displaying the largest amplitude. (b) 795 nm SAS, 780-795 nm SAS, and 780-795 nm DOT-MTS: The blue line illustrates the 795 nm SAS, which includes two resonance peaks and one crossover peak. The green line represents the 780-795 nm SAS, featuring two resonance peaks and four velocity transfer peaks. The red line shows the 780-795 nm DOT-MTS, with the 5$^{2}S_{1/2}$ F=2 $\rightarrow$ 5$^{2}P_{1/2}$ F=1 transition displaying the largest amplitude. (c) Signal gradient of 780 nm MTS and 780-795 nm DOT-MTS at various modulation frequencies.
	}
	\label{spectroscopy}
\end{figure*}

The schematic of the experimental setup and the energy level structure of rubidium atoms are shown in Fig.~\ref{setup}. The 780 nm and 795 nm laser beams are generated from two external-cavity diode lasers (ECDL 780 and ECDL 795), respectively followed by an isolator to avoid the optical feedback. Firstly we employ the common MTS to lock the 780 nm laser to the rubidium D2 line, where the 780 nm laser is divided into two (strong pump and weak probe) beams by a half wave plate and a polarization beam splitter (PBS).  The pump light (0.5 mW) passes through a $^{87}\text{Rb}$ vapor cell and is modulated by an EOM  at the frequency of 6.5 MHz and the probe light (0.1 mw) counter-propagates with the pump. The ratio of pump to probe power can be adjusted precisely through two half-wavelength plates. The direct current (DC) signal of the first photoelectric detector (PD1) is used to derive the SAS of 780 nm, via scanning the 780 nm laser. As shown in Fig.~\ref{spectroscopy}(a), the SAS of the $^{87}\text{Rb}$ D2 line contains three resonance peaks and three crossover peaks. The alternating current (AC) signal after amplification is mixed with the demodulation signal by a laser servo to derive the 780 nm MTS  [see Fig.~\ref{spectroscopy}(a)]. Afterwards the laser servo provides a first proportional integral differential feedback to stabilize the 780 nm laser to the 5$^{2}$$S_{1/2}$ F=2 $\rightarrow$ 5$^{2}$$P_{3/2}$ F=3 transition, which shows the steepest slope in MTS, in $^{87}\text{Rb}$ atoms.

The 795 nm laser is also divided into two beams. One is used to derive the SAS of 795 nm, which features two resonance peaks and one crossover peak [see Fig.~\ref{spectroscopy}(b)] for calibration. The other passes through the M1 mirror and coincident with the 780 nm probe light. M1 and M2 are dichroic mirrors to separate 780 nm and 795 nm lasers. Since the 780 nm laser has been stabilized to the 5$^{2}$$S_{1/2}$F=2 $\rightarrow$ 5$^{2}$$P_{3/2}$F=3 transition, it couples with the 795 nm probe light through atoms with zero-velocity in the ground state F=2. As a result, the modulation on the 780 nm pump light is transferred to the 795 nm probe light. After passing through the M2 mirror, the 795 nm probe light is incident on PD2. The 780-795 nm SAS  reveals four velocity transfer peaks that are caused by the 780 nm pump light and two by the 780 nm probe light\cite{velocity}, alongside the standard saturated absorption peaks [see Fig.~\ref{spectroscopy}(b)].

The amplified AC signal is mixed with the demodulation signal to derive the  780-795 nm DOT-MTS,  which is used to lock the 795 nm laser to the 5$^{2}$$S_{1/2}$F=2 $\rightarrow$ 5$^{2}$$P_{1/2}$F=1 transition in $^{87}\text{Rb}$ atoms. 
Thus, the initial stabilization of the 780 nm laser via MTS is essential to the overall stabilization process. As shown in Fig.~\ref{spectroscopy}(b), the DOT-MTS includes two dispersion curves corresponding to the transitions 5$^{2}$$S_{1/2}$F=2 $\rightarrow$ 5$^{2}$$P_{1/2}$F=1 and F=2. Under identical conditions including the atomic ensemble, modulation frequency $\delta$, and modulation depth, the amplitude of the 780 nm MTS and the DOT-MTS are essentially the same. However, the linewidth of the DOT-MTS is twice as broad as that of the single frequency 780 nm MTS. This results in a twofold difference in their signal gradients, consistent with the theoretical explanations provided (see below). Figure 2(c) illustrates the signal gradients of the 780 nm MTS and DOT-MTS across various $\delta$, both exhibiting maxima at 6.5 MHz. The corresponding spectral linewidth for the 5$^{2}$$S_{1/2}$F=2 $\rightarrow$ 5$^{2}$$P_{3/2}$F=3 transition is approximately 25 MHz.

\begin{figure}[t]
	\begin{center}
		\includegraphics[width=8.8cm,height=4.5cm]{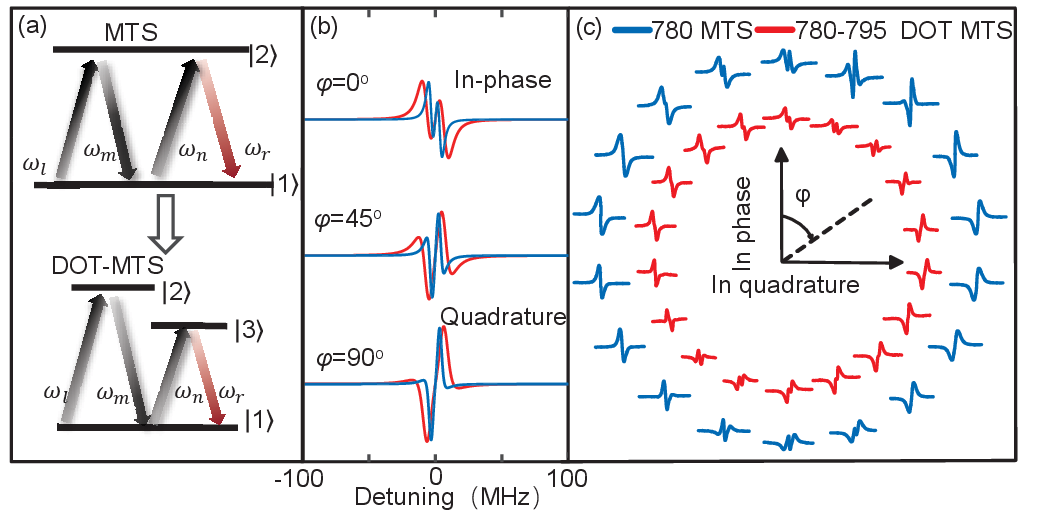}
	\end{center}
	\vspace{-6mm}
	\caption{\label{phase}
		(a) Modulation transfer process for MTS and DOT-MTS and lineshape. 
		(b) Comparison of lineshapes for 780 nm MTS and 780 nm-795 nm DOT-MTS, resulting from theoretical calculations, including in-phase, mixed components, and quadrature phases. 
		(c) Phase change of 780 nm MTS and 780 nm-795 nm DOT-MTS observed experimentally, following a clockwise direction.}
\end{figure}

We model the DOT-MTS using the four-wave mixing in a V-type atomic system. As illustrated in  Fig.~\ref{phase}(a), the ground state and the two excited states of $^{87}\text{Rb}$ are labeled as $|1\rangle$, $|2\rangle$ and $|3\rangle$,corresponding to states
5$^{2}$$S_{1/2}$F=2, 5$^{2}$$P_{3/2}$F=2, 5$^{2}$$P_{1/2}$F=1, respectively. The Lindblad master equation describes the dynamics of V-system:
\begin{align}
	\dot{\hat{\rho }}=\frac{1}{\text{i}\hbar }[{{\hat{H}}_{0}}+\hat{V},\hat{\rho }]+{{\gamma }_{2}}\mathcal{D}[{{\hat{c}}_{2}}]\hat{\rho }+{{\gamma }_{3}}\mathcal{D}[{{\hat{c}}_{3}}]\hat{\rho },    
\end{align}
 where $\hat \rho$ is the density matrix operator, $\hat V$ is the interaction Hamiltonian, $\hat c_2=|1\rangle\langle 2|$, $\hat c_3=|1\rangle\langle 3|$, $\mathcal{D}[\hat{c}]\hat{\rho}=\hat{c}\hat{\rho}\hat{c}^{\dagger}-\frac{1}{2}(\hat{c}^{\dagger}\hat{c}\hat{\rho}+\hat{\rho}\hat{c}^{\dagger}\hat{c})$ is the Lindblad superoperator, $\gamma_2$ and $\gamma_3$ are relaxation rates of excited levels $|2\rangle$ and $|3\rangle$. The Hamiltonian of the unperturbed atomic system is given by $H_0=\hbar\omega_{21} \hat{c}_2^{\dagger}\hat{c}_2+\hbar \omega_{31}\hat{c}_3^{\dagger} \hat{c}_3$, where the $\omega_{ij}$ is the transition frequency between $|i\rangle$ and $|j\rangle$.
\par

In the DOT-MTS, the pump beam for the transition $|1\rangle \to |2\rangle$  is modulated, while the counter-propagating probe beam is detected over the transition $|1\rangle \to |3\rangle$. Here we consider only the carrier and the first-order sidebands of the pump beam due to the small modulation index.  Consequently, the pump beam is described by $E_{pump} =\frac{E_0}{2}[J_0(\kappa)\mathrm{e}^{\mathrm{i}k_cz-\mathrm{i}\omega_ct}+J_1(\kappa)\mathrm{e}^{\mathrm{i}k_cz-\mathrm{i}(\omega_c+\delta)t}\nonumber-J_1(\kappa)\mathrm{e}^{\mathrm{i}k_cz-\mathrm{i}(\omega_c-\delta)t}+c.c]$,
where $E_0$ is the amplitude, $\kappa$ is the modulation index, $J_n(\kappa)$ is the n-order Bessel function, $\omega_c$ is the frequency of the carrier, $k_c=\omega_c/c$ is the wave vector of the carrier, and $\delta$ is the modulation frequency. Assuming $\delta$ is small compared to $\omega_c$, the wave vectors of the sidebands are approximately equal to that of the carrier.
 Similarly, the counter-propagating probe beam can be described by
 $E_{probe} =\frac{E_p}{2}[\mathrm{e}^{-\mathrm{i}(\omega_p t+k_pz)}+c.c]$,
where $E_p$, $k_p$ and $\omega_p$ are the amplitude, wave vector and frequency of the probe beam respectively. Therefore, the interaction Hamiltonian can be given by
\begin{align}
		\hat{V} =&-\frac{\hbar}{2}\big[(\Omega_c\mathrm{e}^{\mathrm{i}k_c z-\mathrm{i}\omega_c t}+\Omega_s\mathrm{e}^{\mathrm{i}k_c z-\mathrm{i}(\omega_c+\delta)t}\nonumber\\
		&-\Omega_s \mathrm{e}^{\mathrm{i}k_cz-\mathrm{i}(\omega_c-\delta)t})(  \hat{c}_2+\hat{c}_2^{\dagger})\nonumber\\
		&-\Omega_p\mathrm{e}^{-\mathrm{i}k_pz-\mathrm{i}\omega_p t}(  \hat{c}_3+\hat{c}_3^{\dagger})\big]+h.c.\ .
\end{align}
Here $\Omega_k=\mu_{ij}E_k/\hbar$ is the Rabi frequency for $k=c$ (carrier), $s$ (sidebands) and $p$ (probe). $\mu_{ij}$ is the matrix element of the dipole operator related to $|i\rangle$ and $|j\rangle$ and is assumed to be real. 
\par
Then, we use the third-order perturbation theory to describe the four-wave mixing and modulation transfer process. In four-wave mixing, new photons carrying modulation information are re-emitted along the same direction as the probe beam, and their generation is proportional to the induced nonlinear macroscopic polarization of the system. By applying the rotating wave approximation and solving for the system's steady state, we obtain the third-order perturbation density matrix element $\rho_{13}^{(3)}$ related to the macroscopic polarization, which in the laboratory frame can be written as
$P^{(3)}=\mu_{31}\int_{\infty}^{\infty}\rho_{13}^{(3)}(v)f(v)\mathrm{d}v+c.c$,
where $f(v)={\mathrm{e}^{-v^2/u^2}}/{u\sqrt{\pi}}$ is the Maxwell velocity distribution and $u$ is the most probable speed of atoms.

\begin{figure*}[t]
	\includegraphics[width=17cm,height=7cm]{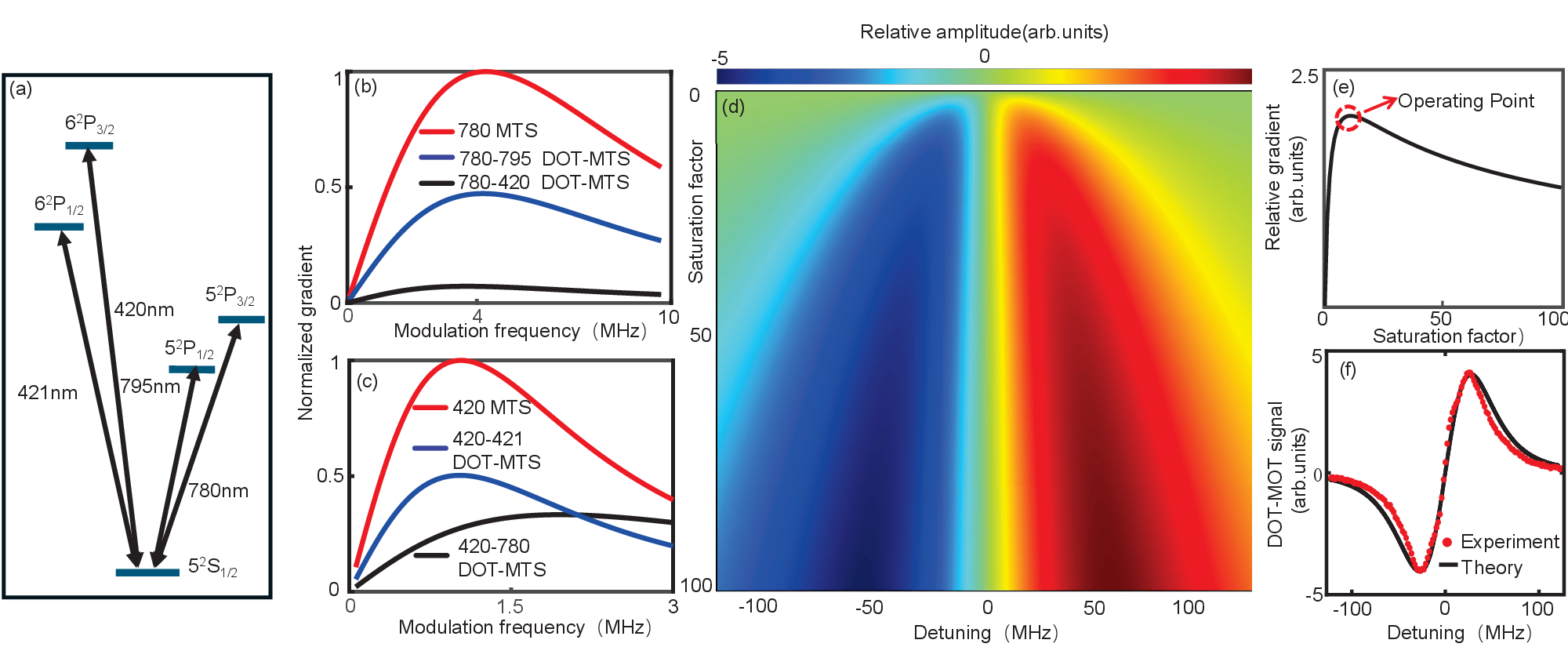}
	\caption{The theoretical analysis examines the DOT-MTS of different transitions and saturation effects for the 780-795 nm DOT-MTS.
		(a) The relevant energy level diagrams.
		(b),(c) The slope gradients of DOT-MTS between different transitions as a function of $\delta$. The legend specifies wavelengths, with the first number denoting the pump beam's wavelength, and the second number the probe beam's wavelength.
		(d) The lineshape when considering saturation effects for the 780-795 nm DOT-MTS. (e) The slope gradient, considering saturation, peaks at a saturation parameter of approximately fifteen, selected as the experimental operating point. (f) The analysis features the DOT-MTS at the optimal saturation parameter, including experimental data and theoretical fitting.
	}
	\label{theory}
\end{figure*}

For an optically thin sample, the re-emitted light fields in the four-wave mixing process can be described as
${E}_{re}=-\mathrm{i}\frac{k_p L}{2\varepsilon_0}P^{(3)}$,
where $k_p$ is the wave vector of the probe beam, $L$ is the length of the vapor cell, $\varepsilon_0$ is the vacuum permittivity. The frequency and wave vector of the new laser generated by four-wave mixing satisfy the momentum and energy conservation conditions:
$\omega_l-\omega_m+\omega_n=\omega_r, \
\boldsymbol{k}_l-\boldsymbol{k}_m+\boldsymbol{k}_n=\boldsymbol{k}_r $,
where the subscript symbols are shown in Fig.~\ref{phase}(a). Table (\ref{tab1}) summarizes  the four combinations of frequency and wave vector conditions, as well as
the detuning of the three-photon resonance peaks relative to the one-photon resonance frequency $\omega_0$ of the probe transition, similar to the situation in MTS described in the literature[\cite{Theory1,Theory2}].

\begin{table}[b]
	\caption{\label{tab1}%
		The frequency combinations generating sidebands in the probe light during four-wave mixing.
	}
	\begin{ruledtabular}
		\begin{tabular}{lcccl} 
			$\omega_l, k_l$ &  $\omega_m, k_m$ & $\omega_n, k_n$ & $\omega_r, k_r$  & $\omega_p-\omega_0$ \\
			\colrule \rule{0pt}{1.2em}%
			$ \omega_c + \delta, k_c$ & $\omega_c, k_c$ & $\omega_p, -k_p$ & $\omega_p + \delta, -k_p$ & $+\delta$ \\
			$\omega_c - \delta, k_c$ & $\omega_c, k_c$ & $\omega_p, -k_p$ & $\omega_p - \delta, -k_p$ & $-\delta-\delta\cdot({\lambda_1/\lambda_2})$  \\
			$\omega_c, k_c$ & $\omega_c + \delta, k_c$ & $\omega_p, -k_p$ & $\omega_p - \delta, -k_p$ & $-\delta$   \\
			$\omega_c, k_c$ & $\omega_c - \delta, k_c$ & $\omega_p, -k_p$ & $\omega_p + \delta, -k_p$&$+\delta+\delta\cdot({\lambda_1/\lambda_2})$\\
		\end{tabular}
	\end{ruledtabular}
\end{table}

Following the theoretical framework, the demodulated spectrum of the DOT-MTS is shown in Fig.~\ref{phase}(b), featuring three distinct spectra: in-phase, mixed components, and quadrature phases. The red line represents the DOT-MTS line shape, while the blue curve shows the normal 780 nm MTS signal for comparison. Phase changes in MTS and DOT-MTS patterns experimentally match theoretical results, as shown in Fig.~\ref{phase}(c). We find that the DOT-MTS linewidth is approximately twice that of the MTS, despite nearly identical signal amplitudes, which is consistent with the phenomenon observed in hole burning \cite{hole} and our experiment. The theoretical $\delta$ is 4.2 MHz (equivalent to 0.7 $\gamma_2/2\pi$), differing from the experimental 6.5 MHz because saturation effects are not considered in the theoretical simulations.

Moreover, numerical calculations indicate that DOT-MTS is primarily constrained by probe relaxation and the ratio of pump to probe wavelengths. Corresponding energy level diagrams are also provided as shown in  Fig.~\ref{theory}(a). The normalized zero crossing gradients of DOT-MTS for various transitions are depicted as a function of $\delta$ in Fig.~\ref{theory}(b),(c). Here, the gradient of the 780 nm MTS is used for normalization purposes. For all depicted curves, the pump laser intensity is consistently set at one-tenth of the saturation intensity. For transitions with similar wavelengths and emission rates, such as 780–795 nm or 420–421 nm, the DOT-MTS gradient is approximately half that of the single quantum transition MTS, with twice the linewidth. When the spontaneous emission rates of two optical transitions differ significantly, modulation transfer efficiency at a constant pump Rabi frequency is limited by the probe transition's emission rate. In our study, the 780-420 nm DOT-MTS is limited by the 420 nm transition's radiation rate. With significant wavelength differences between two optical transitions, the smaller wavelength ratio between the pump and probe transitions in a four-wave mixing process causes the two external three-photon resonance peaks (see Table (\ref{tab1})) to converge inward. This convergence interferes with the error signal, reducing the DOT-MTS gradient. Our study on 420-780 nm DOT-MTS exemplifies this effect, primarily due to the Doppler shift in thermal optical transitions.


We also investigate the lineshape, considering saturation effects, using numerical methods for the 780–795 nm DOT-MTS\cite{bloch1982dispersive}. The effects of spectral broadening and increased amplitude due to saturation are depicted in Fig.~\ref{theory}(d), with each case normalized to when the saturation parameter equals one. The results illustrate notable changes in the lineshape: as the saturation parameter increases, both the broadening of the frequency and the amplitude of the line shape are enhanced. Fig.~\ref{theory}(e) illustrates the slope gradient, taking into account saturation effects, and reveals a maximum at a saturation parameter of approximately fifteen, which is selected as the experimental operating point. The corresponding DOT-MTS experimental and theoretical results, depicted in Fig.~\ref{theory}(f), emphasize this peak gradient and further confirm the validity of the theoretical model.

\begin{figure}[t]
	\begin{center}
		\includegraphics[width=7.6cm,height=8cm]{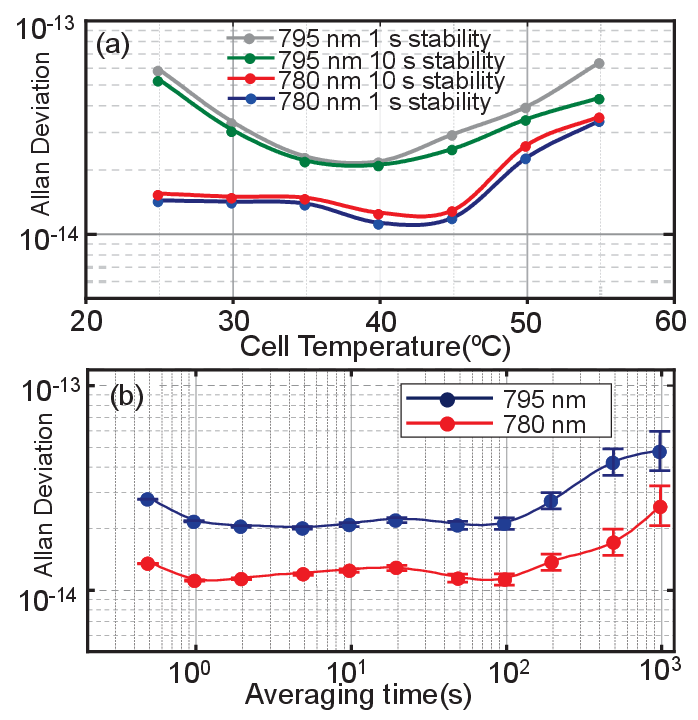}
	\end{center}
	\vspace{-6mm}
	\caption{(a) Instability at different temperatures: The 1 s and 10 s instability of 780 nm laser and 795 nm laser as a function of varying temperatures. (b) Long-term instability at optimal temperature: the long-term instability of the 780 nm and 795 nm lasers at the optimal temperature of 40 $^{\circ}$C.}
	\label{long stabality}
\end{figure}

Through theoretical analysis, DOT-MTS can be employed for multi-wavelength locking, thereby establishing a multi-wavelength frequency standard within the same atomic ensemble. Experimentally validating this, we applied the 780 nm MTS and the DOT-MTS to stabilize the 780 nm and 795 nm lasers, respectively. We investigated the instability of both lasers using the residual error signal, with the Allan deviation results presented in Fig.~\ref{long stabality}. Fig.\ref{long stabality}(a) displays the 1 s and 10 s instability for both lasers at varying temperatures of the atomic cell. At the optimal temperature of 40 $^{\circ}$C, the instability of the 780 nm laser is $1.13 \times 10^{-14}$ at 1 s, and the instability of the 795 nm laser is 
$2.2 \times 10^{-14}$ at 1 s. The long-term instability is depicted in Fig.\ref{long stabality}(b), where the instability of the 795 nm laser is found to be twice as large as that of the 780 nm laser. This discrepancy aligns with the difference in the slope of their error signals. Consequently, the 780 nm and 795 nm lasers are both stabilized to the same atomic ensemble through the utilization of the DOT-MTS, demonstrating excellent in-loop locking precision. 
The self-estimated stability confirms the dual-wavelength optical frequency standard’s ability to “closely track” the atomic frequency and verifies its coherence. However, it typically degrades by an order of magnitude compared to that measured via heterodyne methods between two identical systems \cite{ac3}. The stability of our system, constrained by the inherent frequency fluctuations of the ensemble, can be optimized by extending the approach to atomic beam or cold atom systems, thereby establishing a more precise multi-wavelength standard in the future.

This study successfully demonstrates DOT-MTS on the $^{87}$Rb
D1 and D2 lines, with both 780 nm and 795 nm lasers frequency-stabilized to a single atomic cell via DOT-MTS, achieving high locking precision. We also developed a general expression for DOT-MTS based on a V-type energy level configuration. This theoretical model not only supports our experimental results but also enhances the understanding of the modulation transfer mechanisms. The thermal atomic ensemble may experience frequency fluctuations and long-term drift due to atomic collisions and motion, yet the atomic resonance frequency shifts due to external factors such as temperature and magnetic fields, the locking points of both wavelengths shift in tandem, ensuring high coherence between the dual-wavelength frequency standards. Extending this approach to other atomic systems, such as the $^{39}$K D1 (770 nm) and D2 (766 nm) lines, enables the realization of an integrated optical-THz frequency reference. Similarly, for  $^{87}$Rb D1 line (795 nm) and ground-state microwave transition, it is possible to achieve an integrated optical-microwave frequency reference.

Looking ahead, future work could focus on extending the DOT-MTS to a multi-frequency scheme, incorporating V-type,  $\Lambda$-type, and ladder-type configurations in various atomic structures. Next, we will focus on the iodine molecule, with its rich spectral features, holds great potential for developing multi-wavelength standards that can enhance precision measurements. These advancements would expand applications beyond optical clocks and length metrology, establishing a new frontier in multi-wavelength optical frequency standards realized with a single quantum atomic ensemble and significantly advancing laser stabilization, with far-reaching implications for quantum metrology.

This work was funded by the Beijing Nova Program (No. 20240484696), the INNOVATION Program for Quantum Science and Technology (No. 2021ZD0303202), and the Wenzhou Major Science and Technology Innovation Key Project (No. ZG2020046).
\bibliography{DOMTS}

\begin{thebibliography}{53}%
\makeatletter
\providecommand \@ifxundefined [1]{%
 \@ifx{#1\undefined}
}%
\providecommand \@ifnum [1]{%
 \ifnum #1\expandafter \@firstoftwo
 \else \expandafter \@secondoftwo
 \fi
}%
\providecommand \@ifx [1]{%
 \ifx #1\expandafter \@firstoftwo
 \else \expandafter \@secondoftwo
 \fi
}%
\providecommand \natexlab [1]{#1}%
\providecommand \enquote  [1]{``#1''}%
\providecommand \bibnamefont  [1]{#1}%
\providecommand \bibfnamefont [1]{#1}%
\providecommand \citenamefont [1]{#1}%
\providecommand \href@noop [0]{\@secondoftwo}%
\providecommand \href [0]{\begingroup \@sanitize@url \@href}%
\providecommand \@href[1]{\@@startlink{#1}\@@href}%
\providecommand \@@href[1]{\endgroup#1\@@endlink}%
\providecommand \@sanitize@url [0]{\catcode `\\12\catcode `\$12\catcode
  `\&12\catcode `\#12\catcode `\^12\catcode `\_12\catcode `\%12\relax}%
\providecommand \@@startlink[1]{}%
\providecommand \@@endlink[0]{}%
\providecommand \url  [0]{\begingroup\@sanitize@url \@url }%
\providecommand \@url [1]{\endgroup\@href {#1}{\urlprefix }}%
\providecommand \urlprefix  [0]{URL }%
\providecommand \Eprint [0]{\href }%
\providecommand \doibase [0]{http://dx.doi.org/}%
\providecommand \selectlanguage [0]{\@gobble}%
\providecommand \bibinfo  [0]{\@secondoftwo}%
\providecommand \bibfield  [0]{\@secondoftwo}%
\providecommand \translation [1]{[#1]}%
\providecommand \BibitemOpen [0]{}%
\providecommand \bibitemStop [0]{}%
\providecommand \bibitemNoStop [0]{.\EOS\space}%
\providecommand \EOS [0]{\spacefactor3000\relax}%
\providecommand \BibitemShut  [1]{\csname bibitem#1\endcsname}%
\let\auto@bib@innerbib\@empty
\bibitem [{\citenamefont {Haroche}\ and\ \citenamefont
  {Hartmann}(1972)}]{SAS1}%
  \BibitemOpen
  \bibfield  {author} {\bibinfo {author} {\bibfnamefont {S.}~\bibnamefont
  {Haroche}}\ and\ \bibinfo {author} {\bibfnamefont {F.}~\bibnamefont
  {Hartmann}},\ }\href {\doibase 10.1103/PhysRevA.6.1280} {\bibfield  {journal}
  {\bibinfo  {journal} {Phys. Rev. A}\ }\textbf {\bibinfo {volume} {6}},\
  \bibinfo {pages} {1280} (\bibinfo {year} {1972})}\BibitemShut {NoStop}%
\bibitem [{\citenamefont {Akulshin}\ \emph {et~al.}(1990)\citenamefont
  {Akulshin}, \citenamefont {Sautenkov}, \citenamefont {Velichansky},
  \citenamefont {Zibrov},\ and\ \citenamefont {Zverkov}}]{SAS2}%
  \BibitemOpen
  \bibfield  {author} {\bibinfo {author} {\bibfnamefont {A.}~\bibnamefont
  {Akulshin}}, \bibinfo {author} {\bibfnamefont {V.}~\bibnamefont {Sautenkov}},
  \bibinfo {author} {\bibfnamefont {V.}~\bibnamefont {Velichansky}}, \bibinfo
  {author} {\bibfnamefont {A.}~\bibnamefont {Zibrov}}, \ and\ \bibinfo {author}
  {\bibfnamefont {M.}~\bibnamefont {Zverkov}},\ }\href {\doibase
  https://doi.org/10.1016/0030-4018(90)90094-A} {\bibfield  {journal} {\bibinfo
   {journal} {Optics Communications}\ }\textbf {\bibinfo {volume} {77}},\
  \bibinfo {pages} {295} (\bibinfo {year} {1990})}\BibitemShut {NoStop}%
\bibitem [{\citenamefont {Cahuzac}\ and\ \citenamefont {Vetter}(1975)}]{SAS3}%
  \BibitemOpen
  \bibfield  {author} {\bibinfo {author} {\bibfnamefont {P.}~\bibnamefont
  {Cahuzac}}\ and\ \bibinfo {author} {\bibfnamefont {R.}~\bibnamefont
  {Vetter}},\ }\href {\doibase 10.1103/PhysRevLett.34.1070} {\bibfield
  {journal} {\bibinfo  {journal} {Phys. Rev. Lett.}\ }\textbf {\bibinfo
  {volume} {34}},\ \bibinfo {pages} {1070} (\bibinfo {year}
  {1975})}\BibitemShut {NoStop}%
\bibitem [{\citenamefont {Wieman}\ and\ \citenamefont {H\"ansch}(1976)}]{PS1}%
  \BibitemOpen
  \bibfield  {author} {\bibinfo {author} {\bibfnamefont {C.}~\bibnamefont
  {Wieman}}\ and\ \bibinfo {author} {\bibfnamefont {T.~W.}\ \bibnamefont
  {H\"ansch}},\ }\href {\doibase 10.1103/PhysRevLett.36.1170} {\bibfield
  {journal} {\bibinfo  {journal} {Phys. Rev. Lett.}\ }\textbf {\bibinfo
  {volume} {36}},\ \bibinfo {pages} {1170} (\bibinfo {year}
  {1976})}\BibitemShut {NoStop}%
\bibitem [{\citenamefont {Goldsmith}\ \emph {et~al.}(1978)\citenamefont
  {Goldsmith}, \citenamefont {Weber},\ and\ \citenamefont {H\"ansch}}]{PS2}%
  \BibitemOpen
  \bibfield  {author} {\bibinfo {author} {\bibfnamefont {J.~E.~M.}\
  \bibnamefont {Goldsmith}}, \bibinfo {author} {\bibfnamefont {E.~W.}\
  \bibnamefont {Weber}}, \ and\ \bibinfo {author} {\bibfnamefont {T.~W.}\
  \bibnamefont {H\"ansch}},\ }\href {\doibase 10.1103/PhysRevLett.41.1525}
  {\bibfield  {journal} {\bibinfo  {journal} {Phys. Rev. Lett.}\ }\textbf
  {\bibinfo {volume} {41}},\ \bibinfo {pages} {1525} (\bibinfo {year}
  {1978})}\BibitemShut {NoStop}%
\bibitem [{\citenamefont {Torrance}\ \emph {et~al.}(2016)\citenamefont
  {Torrance}, \citenamefont {Sparkes}, \citenamefont {Turner},\ and\
  \citenamefont {Scholten}}]{PS3}%
  \BibitemOpen
  \bibfield  {author} {\bibinfo {author} {\bibfnamefont {J.~S.}\ \bibnamefont
  {Torrance}}, \bibinfo {author} {\bibfnamefont {B.~M.}\ \bibnamefont
  {Sparkes}}, \bibinfo {author} {\bibfnamefont {L.~D.}\ \bibnamefont {Turner}},
  \ and\ \bibinfo {author} {\bibfnamefont {R.~E.}\ \bibnamefont {Scholten}},\
  }\href {\doibase 10.1364/OE.24.011396} {\bibfield  {journal} {\bibinfo
  {journal} {Opt. Express}\ }\textbf {\bibinfo {volume} {24}},\ \bibinfo
  {pages} {11396} (\bibinfo {year} {2016})}\BibitemShut {NoStop}%
\bibitem [{\citenamefont {Poulsen}\ and\ \citenamefont {Winstrup}(1981)}]{TP1}%
  \BibitemOpen
  \bibfield  {author} {\bibinfo {author} {\bibfnamefont {O.}~\bibnamefont
  {Poulsen}}\ and\ \bibinfo {author} {\bibfnamefont {N.~I.}\ \bibnamefont
  {Winstrup}},\ }\href {\doibase 10.1103/PhysRevLett.47.1522} {\bibfield
  {journal} {\bibinfo  {journal} {Phys. Rev. Lett.}\ }\textbf {\bibinfo
  {volume} {47}},\ \bibinfo {pages} {1522} (\bibinfo {year}
  {1981})}\BibitemShut {NoStop}%
\bibitem [{\citenamefont {Touahri}\ \emph {et~al.}(1997)\citenamefont
  {Touahri}, \citenamefont {Acef}, \citenamefont {Clairon}, \citenamefont
  {Zondy}, \citenamefont {Felder}, \citenamefont {Hilico}, \citenamefont {{de
  Beauvoir}}, \citenamefont {Biraben},\ and\ \citenamefont {Nez}}]{TP2}%
  \BibitemOpen
  \bibfield  {author} {\bibinfo {author} {\bibfnamefont {D.}~\bibnamefont
  {Touahri}}, \bibinfo {author} {\bibfnamefont {O.}~\bibnamefont {Acef}},
  \bibinfo {author} {\bibfnamefont {A.}~\bibnamefont {Clairon}}, \bibinfo
  {author} {\bibfnamefont {J.-J.}\ \bibnamefont {Zondy}}, \bibinfo {author}
  {\bibfnamefont {R.}~\bibnamefont {Felder}}, \bibinfo {author} {\bibfnamefont
  {L.}~\bibnamefont {Hilico}}, \bibinfo {author} {\bibfnamefont
  {B.}~\bibnamefont {{de Beauvoir}}}, \bibinfo {author} {\bibfnamefont
  {F.}~\bibnamefont {Biraben}}, \ and\ \bibinfo {author} {\bibfnamefont
  {F.}~\bibnamefont {Nez}},\ }\href {\doibase
  https://doi.org/10.1016/S0030-4018(96)00471-3} {\bibfield  {journal}
  {\bibinfo  {journal} {Optics Communications}\ }\textbf {\bibinfo {volume}
  {133}},\ \bibinfo {pages} {471} (\bibinfo {year} {1997})}\BibitemShut
  {NoStop}%
\bibitem [{\citenamefont {Tai}\ \emph {et~al.}(1989)\citenamefont {Tai},
  \citenamefont {Mysyrowicz}, \citenamefont {Fischer}, \citenamefont
  {Slusher},\ and\ \citenamefont {Cho}}]{TP3}%
  \BibitemOpen
  \bibfield  {author} {\bibinfo {author} {\bibfnamefont {K.}~\bibnamefont
  {Tai}}, \bibinfo {author} {\bibfnamefont {A.}~\bibnamefont {Mysyrowicz}},
  \bibinfo {author} {\bibfnamefont {R.~J.}\ \bibnamefont {Fischer}}, \bibinfo
  {author} {\bibfnamefont {R.~E.}\ \bibnamefont {Slusher}}, \ and\ \bibinfo
  {author} {\bibfnamefont {A.~Y.}\ \bibnamefont {Cho}},\ }\href {\doibase
  10.1103/PhysRevLett.62.1784} {\bibfield  {journal} {\bibinfo  {journal}
  {Phys. Rev. Lett.}\ }\textbf {\bibinfo {volume} {62}},\ \bibinfo {pages}
  {1784} (\bibinfo {year} {1989})}\BibitemShut {NoStop}%
\bibitem [{\citenamefont {Martin}\ \emph
  {et~al.}(2018{\natexlab{a}})\citenamefont {Martin}, \citenamefont {Phelps},
  \citenamefont {Lemke}, \citenamefont {Bigelow}, \citenamefont {Stuhl},
  \citenamefont {Wojcik}, \citenamefont {Holt}, \citenamefont {Coddington},
  \citenamefont {Bishop},\ and\ \citenamefont {Burke}}]{TP4}%
  \BibitemOpen
  \bibfield  {author} {\bibinfo {author} {\bibfnamefont {K.~W.}\ \bibnamefont
  {Martin}}, \bibinfo {author} {\bibfnamefont {G.}~\bibnamefont {Phelps}},
  \bibinfo {author} {\bibfnamefont {N.~D.}\ \bibnamefont {Lemke}}, \bibinfo
  {author} {\bibfnamefont {M.~S.}\ \bibnamefont {Bigelow}}, \bibinfo {author}
  {\bibfnamefont {B.}~\bibnamefont {Stuhl}}, \bibinfo {author} {\bibfnamefont
  {M.}~\bibnamefont {Wojcik}}, \bibinfo {author} {\bibfnamefont
  {M.}~\bibnamefont {Holt}}, \bibinfo {author} {\bibfnamefont {I.}~\bibnamefont
  {Coddington}}, \bibinfo {author} {\bibfnamefont {M.~W.}\ \bibnamefont
  {Bishop}}, \ and\ \bibinfo {author} {\bibfnamefont {J.~H.}\ \bibnamefont
  {Burke}},\ }\href {\doibase 10.1103/PhysRevApplied.9.014019} {\bibfield
  {journal} {\bibinfo  {journal} {Phys. Rev. Appl.}\ }\textbf {\bibinfo
  {volume} {9}},\ \bibinfo {pages} {014019} (\bibinfo {year}
  {2018}{\natexlab{a}})}\BibitemShut {NoStop}%
\bibitem [{\citenamefont {Brazhnikov}\ \emph {et~al.}(2019)\citenamefont
  {Brazhnikov}, \citenamefont {Petersen}, \citenamefont {Coget}, \citenamefont
  {Passilly}, \citenamefont {Maurice}, \citenamefont {Gorecki},\ and\
  \citenamefont {Boudot}}]{DFSDS1}%
  \BibitemOpen
  \bibfield  {author} {\bibinfo {author} {\bibfnamefont {D.}~\bibnamefont
  {Brazhnikov}}, \bibinfo {author} {\bibfnamefont {M.}~\bibnamefont
  {Petersen}}, \bibinfo {author} {\bibfnamefont {G.}~\bibnamefont {Coget}},
  \bibinfo {author} {\bibfnamefont {N.}~\bibnamefont {Passilly}}, \bibinfo
  {author} {\bibfnamefont {V.}~\bibnamefont {Maurice}}, \bibinfo {author}
  {\bibfnamefont {C.}~\bibnamefont {Gorecki}}, \ and\ \bibinfo {author}
  {\bibfnamefont {R.}~\bibnamefont {Boudot}},\ }\href {\doibase
  10.1103/PhysRevA.99.062508} {\bibfield  {journal} {\bibinfo  {journal} {Phys.
  Rev. A}\ }\textbf {\bibinfo {volume} {99}},\ \bibinfo {pages} {062508}
  (\bibinfo {year} {2019})}\BibitemShut {NoStop}%
\bibitem [{\citenamefont {Gusching}\ \emph {et~al.}(2021)\citenamefont
  {Gusching}, \citenamefont {Petersen}, \citenamefont {Passilly}, \citenamefont
  {Brazhnikov}, \citenamefont {Hafiz},\ and\ \citenamefont {Boudot}}]{DFSDS2}%
  \BibitemOpen
  \bibfield  {author} {\bibinfo {author} {\bibfnamefont {A.}~\bibnamefont
  {Gusching}}, \bibinfo {author} {\bibfnamefont {M.}~\bibnamefont {Petersen}},
  \bibinfo {author} {\bibfnamefont {N.}~\bibnamefont {Passilly}}, \bibinfo
  {author} {\bibfnamefont {D.}~\bibnamefont {Brazhnikov}}, \bibinfo {author}
  {\bibfnamefont {M.~A.}\ \bibnamefont {Hafiz}}, \ and\ \bibinfo {author}
  {\bibfnamefont {R.}~\bibnamefont {Boudot}},\ }\href {\doibase
  10.1364/JOSAB.438111} {\bibfield  {journal} {\bibinfo  {journal} {J. Opt.
  Soc. Am. B}\ }\textbf {\bibinfo {volume} {38}},\ \bibinfo {pages} {3254}
  (\bibinfo {year} {2021})}\BibitemShut {NoStop}%
\bibitem [{\citenamefont {Gusching}\ \emph {et~al.}(2023)\citenamefont
  {Gusching}, \citenamefont {Millo}, \citenamefont {Ryger}, \citenamefont
  {Vicarini}, \citenamefont {Hafiz}, \citenamefont {Passilly},\ and\
  \citenamefont {Boudot}}]{DFSDS3}%
  \BibitemOpen
  \bibfield  {author} {\bibinfo {author} {\bibfnamefont {A.}~\bibnamefont
  {Gusching}}, \bibinfo {author} {\bibfnamefont {J.}~\bibnamefont {Millo}},
  \bibinfo {author} {\bibfnamefont {I.}~\bibnamefont {Ryger}}, \bibinfo
  {author} {\bibfnamefont {R.}~\bibnamefont {Vicarini}}, \bibinfo {author}
  {\bibfnamefont {M.~A.}\ \bibnamefont {Hafiz}}, \bibinfo {author}
  {\bibfnamefont {N.}~\bibnamefont {Passilly}}, \ and\ \bibinfo {author}
  {\bibfnamefont {R.}~\bibnamefont {Boudot}},\ }\href {\doibase
  10.1364/OL.485548} {\bibfield  {journal} {\bibinfo  {journal} {Opt. Lett.}\
  }\textbf {\bibinfo {volume} {48}},\ \bibinfo {pages} {1526} (\bibinfo {year}
  {2023})}\BibitemShut {NoStop}%
\bibitem [{\citenamefont {Bjorklund}(1980)}]{FM1}%
  \BibitemOpen
  \bibfield  {author} {\bibinfo {author} {\bibfnamefont {G.~C.}\ \bibnamefont
  {Bjorklund}},\ }\href {\doibase 10.1364/OL.5.000015} {\bibfield  {journal}
  {\bibinfo  {journal} {Opt. Lett.}\ }\textbf {\bibinfo {volume} {5}},\
  \bibinfo {pages} {15} (\bibinfo {year} {1980})}\BibitemShut {NoStop}%
\bibitem [{\citenamefont {Bjorklund}\ and\ \citenamefont
  {Levenson}(1981)}]{FM2}%
  \BibitemOpen
  \bibfield  {author} {\bibinfo {author} {\bibfnamefont {G.~C.}\ \bibnamefont
  {Bjorklund}}\ and\ \bibinfo {author} {\bibfnamefont {M.~D.}\ \bibnamefont
  {Levenson}},\ }\href {\doibase 10.1103/PhysRevA.24.166} {\bibfield  {journal}
  {\bibinfo  {journal} {Phys. Rev. A}\ }\textbf {\bibinfo {volume} {24}},\
  \bibinfo {pages} {166} (\bibinfo {year} {1981})}\BibitemShut {NoStop}%
\bibitem [{\citenamefont {Zhang}\ \emph {et~al.}(2003)\citenamefont {Zhang},
  \citenamefont {Wei}, \citenamefont {Xie},\ and\ \citenamefont {Peng}}]{MTS2}%
  \BibitemOpen
  \bibfield  {author} {\bibinfo {author} {\bibfnamefont {J.}~\bibnamefont
  {Zhang}}, \bibinfo {author} {\bibfnamefont {D.}~\bibnamefont {Wei}}, \bibinfo
  {author} {\bibfnamefont {C.}~\bibnamefont {Xie}}, \ and\ \bibinfo {author}
  {\bibfnamefont {K.}~\bibnamefont {Peng}},\ }\href {\doibase
  10.1364/OE.11.001338} {\bibfield  {journal} {\bibinfo  {journal} {Opt.
  Express}\ }\textbf {\bibinfo {volume} {11}},\ \bibinfo {pages} {1338}
  (\bibinfo {year} {2003})}\BibitemShut {NoStop}%
\bibitem [{\citenamefont {{Ito}}(2000)}]{MTS1}%
  \BibitemOpen
  \bibfield  {author} {\bibinfo {author} {\bibfnamefont {N.}~\bibnamefont
  {{Ito}}},\ }\href {\doibase 10.1063/1.1150672} {\bibfield  {journal}
  {\bibinfo  {journal} {Review of Scientific Instruments}\ }\textbf {\bibinfo
  {volume} {71}},\ \bibinfo {pages} {2655} (\bibinfo {year}
  {2000})}\BibitemShut {NoStop}%
\bibitem [{\citenamefont {Eble}\ and\ \citenamefont
  {Schmidt-Kaler}(2007)}]{MTS3}%
  \BibitemOpen
  \bibfield  {author} {\bibinfo {author} {\bibfnamefont {J.}~\bibnamefont
  {Eble}}\ and\ \bibinfo {author} {\bibfnamefont {F.}~\bibnamefont
  {Schmidt-Kaler}},\ }\href {https://api.semanticscholar.org/CorpusID:10470398}
  {\bibfield  {journal} {\bibinfo  {journal} {Applied Physics B}\ }\textbf
  {\bibinfo {volume} {88}},\ \bibinfo {pages} {563} (\bibinfo {year}
  {2007})}\BibitemShut {NoStop}%
\bibitem [{\citenamefont {McCarron}\ \emph {et~al.}(2008)\citenamefont
  {McCarron}, \citenamefont {King},\ and\ \citenamefont {Cornish}}]{MTS4}%
  \BibitemOpen
  \bibfield  {author} {\bibinfo {author} {\bibfnamefont {D.~J.}\ \bibnamefont
  {McCarron}}, \bibinfo {author} {\bibfnamefont {S.~A.}\ \bibnamefont {King}},
  \ and\ \bibinfo {author} {\bibfnamefont {S.~L.}\ \bibnamefont {Cornish}},\
  }\href {\doibase 10.1088/0957-0233/19/10/105601} {\bibfield  {journal}
  {\bibinfo  {journal} {Measurement Science and Technology}\ }\textbf {\bibinfo
  {volume} {19}},\ \bibinfo {pages} {105601} (\bibinfo {year}
  {2008})}\BibitemShut {NoStop}%
\bibitem [{\citenamefont {Noh}\ \emph {et~al.}(2011)\citenamefont {Noh},
  \citenamefont {Park}, \citenamefont {Li}, \citenamefont {Park},\ and\
  \citenamefont {Cho}}]{MTS5}%
  \BibitemOpen
  \bibfield  {author} {\bibinfo {author} {\bibfnamefont {H.-R.}\ \bibnamefont
  {Noh}}, \bibinfo {author} {\bibfnamefont {S.~E.}\ \bibnamefont {Park}},
  \bibinfo {author} {\bibfnamefont {L.~Z.}\ \bibnamefont {Li}}, \bibinfo
  {author} {\bibfnamefont {J.-D.}\ \bibnamefont {Park}}, \ and\ \bibinfo
  {author} {\bibfnamefont {C.-H.}\ \bibnamefont {Cho}},\ }\href {\doibase
  10.1364/OE.19.023444} {\bibfield  {journal} {\bibinfo  {journal} {Opt.
  Express}\ }\textbf {\bibinfo {volume} {19}},\ \bibinfo {pages} {23444}
  (\bibinfo {year} {2011})}\BibitemShut {NoStop}%
\bibitem [{\citenamefont {Martin}\ \emph
  {et~al.}(2018{\natexlab{b}})\citenamefont {Martin}, \citenamefont {Phelps},
  \citenamefont {Lemke}, \citenamefont {Bigelow}, \citenamefont {Stuhl},
  \citenamefont {Wojcik}, \citenamefont {Holt}, \citenamefont {Coddington},
  \citenamefont {Bishop},\ and\ \citenamefont {Burke}}]{pm1}%
  \BibitemOpen
  \bibfield  {author} {\bibinfo {author} {\bibfnamefont {K.~W.}\ \bibnamefont
  {Martin}}, \bibinfo {author} {\bibfnamefont {G.}~\bibnamefont {Phelps}},
  \bibinfo {author} {\bibfnamefont {N.~D.}\ \bibnamefont {Lemke}}, \bibinfo
  {author} {\bibfnamefont {M.~S.}\ \bibnamefont {Bigelow}}, \bibinfo {author}
  {\bibfnamefont {B.}~\bibnamefont {Stuhl}}, \bibinfo {author} {\bibfnamefont
  {M.}~\bibnamefont {Wojcik}}, \bibinfo {author} {\bibfnamefont
  {M.}~\bibnamefont {Holt}}, \bibinfo {author} {\bibfnamefont {I.}~\bibnamefont
  {Coddington}}, \bibinfo {author} {\bibfnamefont {M.~W.}\ \bibnamefont
  {Bishop}}, \ and\ \bibinfo {author} {\bibfnamefont {J.~H.}\ \bibnamefont
  {Burke}},\ }\href {\doibase 10.1103/PhysRevApplied.9.014019} {\bibfield
  {journal} {\bibinfo  {journal} {Phys. Rev. Appl.}\ }\textbf {\bibinfo
  {volume} {9}},\ \bibinfo {pages} {014019} (\bibinfo {year}
  {2018}{\natexlab{b}})}\BibitemShut {NoStop}%
\bibitem [{\citenamefont {Huntemann}\ \emph {et~al.}(2014)\citenamefont
  {Huntemann}, \citenamefont {Lipphardt}, \citenamefont {Tamm}, \citenamefont
  {Gerginov}, \citenamefont {Weyers},\ and\ \citenamefont {Peik}}]{pm2}%
  \BibitemOpen
  \bibfield  {author} {\bibinfo {author} {\bibfnamefont {N.}~\bibnamefont
  {Huntemann}}, \bibinfo {author} {\bibfnamefont {B.}~\bibnamefont
  {Lipphardt}}, \bibinfo {author} {\bibfnamefont {C.}~\bibnamefont {Tamm}},
  \bibinfo {author} {\bibfnamefont {V.}~\bibnamefont {Gerginov}}, \bibinfo
  {author} {\bibfnamefont {S.}~\bibnamefont {Weyers}}, \ and\ \bibinfo {author}
  {\bibfnamefont {E.}~\bibnamefont {Peik}},\ }\href {\doibase
  10.1103/PhysRevLett.113.210802} {\bibfield  {journal} {\bibinfo  {journal}
  {Phys. Rev. Lett.}\ }\textbf {\bibinfo {volume} {113}},\ \bibinfo {pages}
  {210802} (\bibinfo {year} {2014})}\BibitemShut {NoStop}%
\bibitem [{\citenamefont {Lange}\ \emph {et~al.}(2021)\citenamefont {Lange},
  \citenamefont {Huntemann}, \citenamefont {Rahm}, \citenamefont {Sanner},
  \citenamefont {Shao}, \citenamefont {Lipphardt}, \citenamefont {Tamm},
  \citenamefont {Weyers},\ and\ \citenamefont {Peik}}]{pm3}%
  \BibitemOpen
  \bibfield  {author} {\bibinfo {author} {\bibfnamefont {R.}~\bibnamefont
  {Lange}}, \bibinfo {author} {\bibfnamefont {N.}~\bibnamefont {Huntemann}},
  \bibinfo {author} {\bibfnamefont {J.~M.}\ \bibnamefont {Rahm}}, \bibinfo
  {author} {\bibfnamefont {C.}~\bibnamefont {Sanner}}, \bibinfo {author}
  {\bibfnamefont {H.}~\bibnamefont {Shao}}, \bibinfo {author} {\bibfnamefont
  {B.}~\bibnamefont {Lipphardt}}, \bibinfo {author} {\bibfnamefont
  {C.}~\bibnamefont {Tamm}}, \bibinfo {author} {\bibfnamefont {S.}~\bibnamefont
  {Weyers}}, \ and\ \bibinfo {author} {\bibfnamefont {E.}~\bibnamefont
  {Peik}},\ }\href {\doibase 10.1103/PhysRevLett.126.011102} {\bibfield
  {journal} {\bibinfo  {journal} {Phys. Rev. Lett.}\ }\textbf {\bibinfo
  {volume} {126}},\ \bibinfo {pages} {011102} (\bibinfo {year}
  {2021})}\BibitemShut {NoStop}%
\bibitem [{\citenamefont {Ludlow}\ \emph {et~al.}(2015)\citenamefont {Ludlow},
  \citenamefont {Boyd}, \citenamefont {Ye}, \citenamefont {Peik},\ and\
  \citenamefont {Schmidt}}]{ac2}%
  \BibitemOpen
  \bibfield  {author} {\bibinfo {author} {\bibfnamefont {A.~D.}\ \bibnamefont
  {Ludlow}}, \bibinfo {author} {\bibfnamefont {M.~M.}\ \bibnamefont {Boyd}},
  \bibinfo {author} {\bibfnamefont {J.}~\bibnamefont {Ye}}, \bibinfo {author}
  {\bibfnamefont {E.}~\bibnamefont {Peik}}, \ and\ \bibinfo {author}
  {\bibfnamefont {P.~O.}\ \bibnamefont {Schmidt}},\ }\href {\doibase
  10.1103/RevModPhys.87.637} {\bibfield  {journal} {\bibinfo  {journal} {Rev.
  Mod. Phys.}\ }\textbf {\bibinfo {volume} {87}},\ \bibinfo {pages} {637}
  (\bibinfo {year} {2015})}\BibitemShut {NoStop}%
\bibitem [{\citenamefont {Miao}\ \emph
  {et~al.}(2022{\natexlab{a}})\citenamefont {Miao}, \citenamefont {Shi},
  \citenamefont {Zhang},\ and\ \citenamefont {Chen}}]{ac3}%
  \BibitemOpen
  \bibfield  {author} {\bibinfo {author} {\bibfnamefont {J.}~\bibnamefont
  {Miao}}, \bibinfo {author} {\bibfnamefont {T.}~\bibnamefont {Shi}}, \bibinfo
  {author} {\bibfnamefont {J.}~\bibnamefont {Zhang}}, \ and\ \bibinfo {author}
  {\bibfnamefont {J.}~\bibnamefont {Chen}},\ }\href {\doibase
  10.1103/PhysRevApplied.18.024034} {\bibfield  {journal} {\bibinfo  {journal}
  {Phys. Rev. Appl.}\ }\textbf {\bibinfo {volume} {18}},\ \bibinfo {pages}
  {024034} (\bibinfo {year} {2022}{\natexlab{a}})}\BibitemShut {NoStop}%
\bibitem [{\citenamefont {Newman}\ \emph {et~al.}(2019)\citenamefont {Newman},
  \citenamefont {Maurice}, \citenamefont {Drake}, \citenamefont {Stone},
  \citenamefont {Briles}, \citenamefont {Spencer}, \citenamefont {Fredrick},
  \citenamefont {Li}, \citenamefont {Westly}, \citenamefont {Ilic},
  \citenamefont {Shen}, \citenamefont {Suh}, \citenamefont {Yang},
  \citenamefont {Johnson}, \citenamefont {Johnson}, \citenamefont {Hollberg},
  \citenamefont {Vahala}, \citenamefont {Srinivasan}, \citenamefont {Diddams},
  \citenamefont {Kitching}, \citenamefont {Papp},\ and\ \citenamefont
  {Hummon}}]{ac4}%
  \BibitemOpen
  \bibfield  {author} {\bibinfo {author} {\bibfnamefont {Z.~L.}\ \bibnamefont
  {Newman}}, \bibinfo {author} {\bibfnamefont {V.}~\bibnamefont {Maurice}},
  \bibinfo {author} {\bibfnamefont {T.}~\bibnamefont {Drake}}, \bibinfo
  {author} {\bibfnamefont {J.~R.}\ \bibnamefont {Stone}}, \bibinfo {author}
  {\bibfnamefont {T.~C.}\ \bibnamefont {Briles}}, \bibinfo {author}
  {\bibfnamefont {D.~T.}\ \bibnamefont {Spencer}}, \bibinfo {author}
  {\bibfnamefont {C.}~\bibnamefont {Fredrick}}, \bibinfo {author}
  {\bibfnamefont {Q.}~\bibnamefont {Li}}, \bibinfo {author} {\bibfnamefont
  {D.}~\bibnamefont {Westly}}, \bibinfo {author} {\bibfnamefont {B.~R.}\
  \bibnamefont {Ilic}}, \bibinfo {author} {\bibfnamefont {B.}~\bibnamefont
  {Shen}}, \bibinfo {author} {\bibfnamefont {M.-G.}\ \bibnamefont {Suh}},
  \bibinfo {author} {\bibfnamefont {K.~Y.}\ \bibnamefont {Yang}}, \bibinfo
  {author} {\bibfnamefont {C.}~\bibnamefont {Johnson}}, \bibinfo {author}
  {\bibfnamefont {D.~M.~S.}\ \bibnamefont {Johnson}}, \bibinfo {author}
  {\bibfnamefont {L.}~\bibnamefont {Hollberg}}, \bibinfo {author}
  {\bibfnamefont {K.~J.}\ \bibnamefont {Vahala}}, \bibinfo {author}
  {\bibfnamefont {K.}~\bibnamefont {Srinivasan}}, \bibinfo {author}
  {\bibfnamefont {S.~A.}\ \bibnamefont {Diddams}}, \bibinfo {author}
  {\bibfnamefont {J.}~\bibnamefont {Kitching}}, \bibinfo {author}
  {\bibfnamefont {S.~B.}\ \bibnamefont {Papp}}, \ and\ \bibinfo {author}
  {\bibfnamefont {M.~T.}\ \bibnamefont {Hummon}},\ }\href {\doibase
  10.1364/OPTICA.6.000680} {\bibfield  {journal} {\bibinfo  {journal} {Optica}\
  }\textbf {\bibinfo {volume} {6}},\ \bibinfo {pages} {680} (\bibinfo {year}
  {2019})}\BibitemShut {NoStop}%
\bibitem [{\citenamefont {Fertig}\ and\ \citenamefont {Gibble}(2000)}]{cd1}%
  \BibitemOpen
  \bibfield  {author} {\bibinfo {author} {\bibfnamefont {C.}~\bibnamefont
  {Fertig}}\ and\ \bibinfo {author} {\bibfnamefont {K.}~\bibnamefont
  {Gibble}},\ }\href {\doibase 10.1103/PhysRevLett.85.1622} {\bibfield
  {journal} {\bibinfo  {journal} {Phys. Rev. Lett.}\ }\textbf {\bibinfo
  {volume} {85}},\ \bibinfo {pages} {1622} (\bibinfo {year}
  {2000})}\BibitemShut {NoStop}%
\bibitem [{\citenamefont {Liu}\ \emph {et~al.}(2018)\citenamefont {Liu},
  \citenamefont {L{\"u}}, \citenamefont {Chen}, \citenamefont {Li},
  \citenamefont {Qu}, \citenamefont {Wang}, \citenamefont {Li}, \citenamefont
  {Ren}, \citenamefont {Dong}, \citenamefont {Zhao} \emph {et~al.}}]{cd2}%
  \BibitemOpen
  \bibfield  {author} {\bibinfo {author} {\bibfnamefont {L.}~\bibnamefont
  {Liu}}, \bibinfo {author} {\bibfnamefont {D.-S.}\ \bibnamefont {L{\"u}}},
  \bibinfo {author} {\bibfnamefont {W.-B.}\ \bibnamefont {Chen}}, \bibinfo
  {author} {\bibfnamefont {T.}~\bibnamefont {Li}}, \bibinfo {author}
  {\bibfnamefont {Q.-Z.}\ \bibnamefont {Qu}}, \bibinfo {author} {\bibfnamefont
  {B.}~\bibnamefont {Wang}}, \bibinfo {author} {\bibfnamefont {L.}~\bibnamefont
  {Li}}, \bibinfo {author} {\bibfnamefont {W.}~\bibnamefont {Ren}}, \bibinfo
  {author} {\bibfnamefont {Z.-R.}\ \bibnamefont {Dong}}, \bibinfo {author}
  {\bibfnamefont {J.-B.}\ \bibnamefont {Zhao}},  \emph {et~al.},\ }\href@noop
  {} {\bibfield  {journal} {\bibinfo  {journal} {Nature communications}\
  }\textbf {\bibinfo {volume} {9}},\ \bibinfo {pages} {2760} (\bibinfo {year}
  {2018})}\BibitemShut {NoStop}%
\bibitem [{\citenamefont {Roslund}\ \emph {et~al.}(2024)\citenamefont
  {Roslund}, \citenamefont {Cing{\"o}z}, \citenamefont {Lunden}, \citenamefont
  {Partridge}, \citenamefont {Kowligy}, \citenamefont {Roller}, \citenamefont
  {Sheredy}, \citenamefont {Skulason}, \citenamefont {Song}, \citenamefont
  {Abo-Shaeer} \emph {et~al.}}]{nav1}%
  \BibitemOpen
  \bibfield  {author} {\bibinfo {author} {\bibfnamefont {J.~D.}\ \bibnamefont
  {Roslund}}, \bibinfo {author} {\bibfnamefont {A.}~\bibnamefont {Cing{\"o}z}},
  \bibinfo {author} {\bibfnamefont {W.~D.}\ \bibnamefont {Lunden}}, \bibinfo
  {author} {\bibfnamefont {G.~B.}\ \bibnamefont {Partridge}}, \bibinfo {author}
  {\bibfnamefont {A.~S.}\ \bibnamefont {Kowligy}}, \bibinfo {author}
  {\bibfnamefont {F.}~\bibnamefont {Roller}}, \bibinfo {author} {\bibfnamefont
  {D.~B.}\ \bibnamefont {Sheredy}}, \bibinfo {author} {\bibfnamefont {G.~E.}\
  \bibnamefont {Skulason}}, \bibinfo {author} {\bibfnamefont {J.~P.}\
  \bibnamefont {Song}}, \bibinfo {author} {\bibfnamefont {J.~R.}\ \bibnamefont
  {Abo-Shaeer}},  \emph {et~al.},\ }\href@noop {} {\bibfield  {journal}
  {\bibinfo  {journal} {Nature}\ }\textbf {\bibinfo {volume} {628}},\ \bibinfo
  {pages} {736} (\bibinfo {year} {2024})}\BibitemShut {NoStop}%
\bibitem [{\citenamefont {Ahern}\ \emph {et~al.}(2024)\citenamefont {Ahern},
  \citenamefont {Allison}, \citenamefont {Billington}, \citenamefont {Hébert},
  \citenamefont {Hilton}, \citenamefont {Klantsataya}, \citenamefont {Locke},
  \citenamefont {Luiten}, \citenamefont {Nelligan}, \citenamefont {Offer},
  \citenamefont {Perrella}, \citenamefont {Scholten}, \citenamefont {White},
  \citenamefont {Sparkes}, \citenamefont {Beard}, \citenamefont {Elgin},\ and\
  \citenamefont {Martin}}]{nav2}%
  \BibitemOpen
  \bibfield  {author} {\bibinfo {author} {\bibfnamefont {E.}~\bibnamefont
  {Ahern}}, \bibinfo {author} {\bibfnamefont {J.~W.}\ \bibnamefont {Allison}},
  \bibinfo {author} {\bibfnamefont {C.}~\bibnamefont {Billington}}, \bibinfo
  {author} {\bibfnamefont {N.~B.}\ \bibnamefont {Hébert}}, \bibinfo {author}
  {\bibfnamefont {A.~P.}\ \bibnamefont {Hilton}}, \bibinfo {author}
  {\bibfnamefont {E.}~\bibnamefont {Klantsataya}}, \bibinfo {author}
  {\bibfnamefont {C.}~\bibnamefont {Locke}}, \bibinfo {author} {\bibfnamefont
  {A.~N.}\ \bibnamefont {Luiten}}, \bibinfo {author} {\bibfnamefont
  {M.}~\bibnamefont {Nelligan}}, \bibinfo {author} {\bibfnamefont {R.~F.}\
  \bibnamefont {Offer}}, \bibinfo {author} {\bibfnamefont {C.}~\bibnamefont
  {Perrella}}, \bibinfo {author} {\bibfnamefont {S.~K.}\ \bibnamefont
  {Scholten}}, \bibinfo {author} {\bibfnamefont {B.}~\bibnamefont {White}},
  \bibinfo {author} {\bibfnamefont {B.~M.}\ \bibnamefont {Sparkes}}, \bibinfo
  {author} {\bibfnamefont {R.}~\bibnamefont {Beard}}, \bibinfo {author}
  {\bibfnamefont {J.~D.}\ \bibnamefont {Elgin}}, \ and\ \bibinfo {author}
  {\bibfnamefont {K.~W.}\ \bibnamefont {Martin}},\ }\href
  {https://arxiv.org/abs/2406.03716} {\enquote {\bibinfo {title} {Demonstration
  of a mobile optical clock ensemble at sea},}\ } (\bibinfo {year} {2024}),\
  \Eprint {http://arxiv.org/abs/2406.03716} {arXiv:2406.03716
  [physics.atom-ph]} \BibitemShut {NoStop}%
\bibitem [{\citenamefont {Miao}\ \emph
  {et~al.}(2022{\natexlab{b}})\citenamefont {Miao}, \citenamefont {Shi},
  \citenamefont {Zhang},\ and\ \citenamefont {Chen}}]{com1}%
  \BibitemOpen
  \bibfield  {author} {\bibinfo {author} {\bibfnamefont {J.}~\bibnamefont
  {Miao}}, \bibinfo {author} {\bibfnamefont {T.}~\bibnamefont {Shi}}, \bibinfo
  {author} {\bibfnamefont {J.}~\bibnamefont {Zhang}}, \ and\ \bibinfo {author}
  {\bibfnamefont {J.}~\bibnamefont {Chen}},\ }\href {\doibase
  10.1103/PhysRevApplied.18.024034} {\bibfield  {journal} {\bibinfo  {journal}
  {Phys. Rev. Appl.}\ }\textbf {\bibinfo {volume} {18}},\ \bibinfo {pages}
  {024034} (\bibinfo {year} {2022}{\natexlab{b}})}\BibitemShut {NoStop}%
\bibitem [{\citenamefont {Ye}\ \emph {et~al.}(2001)\citenamefont {Ye},
  \citenamefont {Ma},\ and\ \citenamefont {Hall}}]{I1}%
  \BibitemOpen
  \bibfield  {author} {\bibinfo {author} {\bibfnamefont {J.}~\bibnamefont
  {Ye}}, \bibinfo {author} {\bibfnamefont {L.~S.}\ \bibnamefont {Ma}}, \ and\
  \bibinfo {author} {\bibfnamefont {J.~L.}\ \bibnamefont {Hall}},\ }\href
  {\doibase 10.1103/PhysRevLett.87.270801} {\bibfield  {journal} {\bibinfo
  {journal} {Phys. Rev. Lett.}\ }\textbf {\bibinfo {volume} {87}},\ \bibinfo
  {pages} {270801} (\bibinfo {year} {2001})}\BibitemShut {NoStop}%
\bibitem [{\citenamefont {Ma}\ \emph {et~al.}(1999)\citenamefont {Ma},
  \citenamefont {Ye}, \citenamefont {Dub\'{e}},\ and\ \citenamefont
  {Hall}}]{ace1}%
  \BibitemOpen
  \bibfield  {author} {\bibinfo {author} {\bibfnamefont {L.-S.}\ \bibnamefont
  {Ma}}, \bibinfo {author} {\bibfnamefont {J.}~\bibnamefont {Ye}}, \bibinfo
  {author} {\bibfnamefont {P.}~\bibnamefont {Dub\'{e}}}, \ and\ \bibinfo
  {author} {\bibfnamefont {J.~L.}\ \bibnamefont {Hall}},\ }\href {\doibase
  10.1364/JOSAB.16.002255} {\bibfield  {journal} {\bibinfo  {journal} {J. Opt.
  Soc. Am. B}\ }\textbf {\bibinfo {volume} {16}},\ \bibinfo {pages} {2255}
  (\bibinfo {year} {1999})}\BibitemShut {NoStop}%
\bibitem [{\citenamefont {Sun}\ \emph {et~al.}(2016)\citenamefont {Sun},
  \citenamefont {Zhou}, \citenamefont {Zhou}, \citenamefont {Wang},\ and\
  \citenamefont {Zhan}}]{alkali1}%
  \BibitemOpen
  \bibfield  {author} {\bibinfo {author} {\bibfnamefont {D.}~\bibnamefont
  {Sun}}, \bibinfo {author} {\bibfnamefont {C.}~\bibnamefont {Zhou}}, \bibinfo
  {author} {\bibfnamefont {L.}~\bibnamefont {Zhou}}, \bibinfo {author}
  {\bibfnamefont {J.}~\bibnamefont {Wang}}, \ and\ \bibinfo {author}
  {\bibfnamefont {M.}~\bibnamefont {Zhan}},\ }\href {\doibase
  10.1364/OE.24.010649} {\bibfield  {journal} {\bibinfo  {journal} {Opt.
  Express}\ }\textbf {\bibinfo {volume} {24}},\ \bibinfo {pages} {10649}
  (\bibinfo {year} {2016})}\BibitemShut {NoStop}%
\bibitem [{\citenamefont {Mudarikwa}\ \emph {et~al.}(2012)\citenamefont
  {Mudarikwa}, \citenamefont {Pahwa},\ and\ \citenamefont {Goldwin}}]{alkali2}%
  \BibitemOpen
  \bibfield  {author} {\bibinfo {author} {\bibfnamefont {L.}~\bibnamefont
  {Mudarikwa}}, \bibinfo {author} {\bibfnamefont {K.}~\bibnamefont {Pahwa}}, \
  and\ \bibinfo {author} {\bibfnamefont {J.}~\bibnamefont {Goldwin}},\ }\href
  {\doibase 10.1088/0953-4075/45/6/065002} {\bibfield  {journal} {\bibinfo
  {journal} {Journal of Physics B: Atomic, Molecular and Optical Physics}\
  }\textbf {\bibinfo {volume} {45}},\ \bibinfo {pages} {065002} (\bibinfo
  {year} {2012})}\BibitemShut {NoStop}%
\bibitem [{\citenamefont {Bertinetto}\ \emph {et~al.}(2001)\citenamefont
  {Bertinetto}, \citenamefont {Cordiale}, \citenamefont {Galzerano},\ and\
  \citenamefont {Bava}}]{alkali3}%
  \BibitemOpen
  \bibfield  {author} {\bibinfo {author} {\bibfnamefont {F.}~\bibnamefont
  {Bertinetto}}, \bibinfo {author} {\bibfnamefont {P.}~\bibnamefont
  {Cordiale}}, \bibinfo {author} {\bibfnamefont {G.}~\bibnamefont {Galzerano}},
  \ and\ \bibinfo {author} {\bibfnamefont {E.}~\bibnamefont {Bava}},\ }\href
  {\doibase 10.1109/19.918173} {\bibfield  {journal} {\bibinfo  {journal} {IEEE
  Transactions on Instrumentation and Measurement}\ }\textbf {\bibinfo {volume}
  {50}},\ \bibinfo {pages} {490} (\bibinfo {year} {2001})}\BibitemShut
  {NoStop}%
\bibitem [{\citenamefont {Lee}\ \emph {et~al.}(2023)\citenamefont {Lee},
  \citenamefont {Moon}, \citenamefont {Park}, \citenamefont {Hong},
  \citenamefont {Lee}, \citenamefont {Seo}, \citenamefont {Kwon},\ and\
  \citenamefont {Lee}}]{Best1}%
  \BibitemOpen
  \bibfield  {author} {\bibinfo {author} {\bibfnamefont {S.}~\bibnamefont
  {Lee}}, \bibinfo {author} {\bibfnamefont {G.}~\bibnamefont {Moon}}, \bibinfo
  {author} {\bibfnamefont {S.~E.}\ \bibnamefont {Park}}, \bibinfo {author}
  {\bibfnamefont {H.-G.}\ \bibnamefont {Hong}}, \bibinfo {author}
  {\bibfnamefont {J.~H.}\ \bibnamefont {Lee}}, \bibinfo {author} {\bibfnamefont
  {S.}~\bibnamefont {Seo}}, \bibinfo {author} {\bibfnamefont {T.~Y.}\
  \bibnamefont {Kwon}}, \ and\ \bibinfo {author} {\bibfnamefont {S.-B.}\
  \bibnamefont {Lee}},\ }\href {\doibase 10.1364/OL.480178} {\bibfield
  {journal} {\bibinfo  {journal} {Opt. Lett.}\ }\textbf {\bibinfo {volume}
  {48}},\ \bibinfo {pages} {1020} (\bibinfo {year} {2023})}\BibitemShut
  {NoStop}%
\bibitem [{\citenamefont {Udem}\ \emph {et~al.}(2002)\citenamefont {Udem},
  \citenamefont {Holzwarth},\ and\ \citenamefont {H{\"a}nsch}}]{metro}%
  \BibitemOpen
  \bibfield  {author} {\bibinfo {author} {\bibfnamefont {T.}~\bibnamefont
  {Udem}}, \bibinfo {author} {\bibfnamefont {R.}~\bibnamefont {Holzwarth}}, \
  and\ \bibinfo {author} {\bibfnamefont {T.~W.}\ \bibnamefont {H{\"a}nsch}},\
  }\href@noop {} {\bibfield  {journal} {\bibinfo  {journal} {Nature}\ }\textbf
  {\bibinfo {volume} {416}},\ \bibinfo {pages} {233} (\bibinfo {year}
  {2002})}\BibitemShut {NoStop}%
\bibitem [{\citenamefont {Torres-Company}\ and\ \citenamefont
  {Weiner}(2014)}]{tele1}%
  \BibitemOpen
  \bibfield  {author} {\bibinfo {author} {\bibfnamefont {V.}~\bibnamefont
  {Torres-Company}}\ and\ \bibinfo {author} {\bibfnamefont {A.~M.}\
  \bibnamefont {Weiner}},\ }\href@noop {} {\bibfield  {journal} {\bibinfo
  {journal} {Laser \& Photonics Reviews}\ }\textbf {\bibinfo {volume} {8}},\
  \bibinfo {pages} {368} (\bibinfo {year} {2014})}\BibitemShut {NoStop}%
\bibitem [{\citenamefont {Miller}(2012)}]{tele2}%
  \BibitemOpen
  \bibfield  {author} {\bibinfo {author} {\bibfnamefont {S.}~\bibnamefont
  {Miller}},\ }\href@noop {} {\emph {\bibinfo {title} {Optical fiber
  telecommunications}}}\ (\bibinfo  {publisher} {Elsevier},\ \bibinfo {year}
  {2012})\BibitemShut {NoStop}%
\bibitem [{\citenamefont {T{\'o}bi{\'a}s}\ \emph {et~al.}(2020)\citenamefont
  {T{\'o}bi{\'a}s}, \citenamefont {Furtenbacher}, \citenamefont {Simk{\'o}},
  \citenamefont {Cs{\'a}sz{\'a}r}, \citenamefont {Diouf}, \citenamefont
  {Cozijn}, \citenamefont {Staa}, \citenamefont {Salumbides},\ and\
  \citenamefont {Ubachs}}]{network1}%
  \BibitemOpen
  \bibfield  {author} {\bibinfo {author} {\bibfnamefont {R.}~\bibnamefont
  {T{\'o}bi{\'a}s}}, \bibinfo {author} {\bibfnamefont {T.}~\bibnamefont
  {Furtenbacher}}, \bibinfo {author} {\bibfnamefont {I.}~\bibnamefont
  {Simk{\'o}}}, \bibinfo {author} {\bibfnamefont {A.~G.}\ \bibnamefont
  {Cs{\'a}sz{\'a}r}}, \bibinfo {author} {\bibfnamefont {M.~L.}\ \bibnamefont
  {Diouf}}, \bibinfo {author} {\bibfnamefont {F.~M.}\ \bibnamefont {Cozijn}},
  \bibinfo {author} {\bibfnamefont {J.~M.}\ \bibnamefont {Staa}}, \bibinfo
  {author} {\bibfnamefont {E.~J.}\ \bibnamefont {Salumbides}}, \ and\ \bibinfo
  {author} {\bibfnamefont {W.}~\bibnamefont {Ubachs}},\ }\href@noop {}
  {\bibfield  {journal} {\bibinfo  {journal} {Nature Communications}\ }\textbf
  {\bibinfo {volume} {11}},\ \bibinfo {pages} {1708} (\bibinfo {year}
  {2020})}\BibitemShut {NoStop}%
\bibitem [{\citenamefont {Castrillo}\ \emph {et~al.}(2023)\citenamefont
  {Castrillo}, \citenamefont {Fasci}, \citenamefont {Furtenbacher},
  \citenamefont {D'Agostino}, \citenamefont {Khan}, \citenamefont {Gravina},
  \citenamefont {Gianfrani},\ and\ \citenamefont {Cs{\'a}sz{\'a}r}}]{network2}%
  \BibitemOpen
  \bibfield  {author} {\bibinfo {author} {\bibfnamefont {A.}~\bibnamefont
  {Castrillo}}, \bibinfo {author} {\bibfnamefont {E.}~\bibnamefont {Fasci}},
  \bibinfo {author} {\bibfnamefont {T.}~\bibnamefont {Furtenbacher}}, \bibinfo
  {author} {\bibfnamefont {V.}~\bibnamefont {D'Agostino}}, \bibinfo {author}
  {\bibfnamefont {M.~A.}\ \bibnamefont {Khan}}, \bibinfo {author}
  {\bibfnamefont {S.}~\bibnamefont {Gravina}}, \bibinfo {author} {\bibfnamefont
  {L.}~\bibnamefont {Gianfrani}}, \ and\ \bibinfo {author} {\bibfnamefont
  {A.~G.}\ \bibnamefont {Cs{\'a}sz{\'a}r}},\ }\href@noop {} {\bibfield
  {journal} {\bibinfo  {journal} {Physical Chemistry Chemical Physics}\
  }\textbf {\bibinfo {volume} {25}},\ \bibinfo {pages} {23614} (\bibinfo {year}
  {2023})}\BibitemShut {NoStop}%
\bibitem [{\citenamefont {Raj}\ \emph {et~al.}(1980)\citenamefont {Raj},
  \citenamefont {Bloch}, \citenamefont {Snyder}, \citenamefont {Camy},\ and\
  \citenamefont {Ducloy}}]{Theory1}%
  \BibitemOpen
  \bibfield  {author} {\bibinfo {author} {\bibfnamefont {R.~K.}\ \bibnamefont
  {Raj}}, \bibinfo {author} {\bibfnamefont {D.}~\bibnamefont {Bloch}}, \bibinfo
  {author} {\bibfnamefont {J.~J.}\ \bibnamefont {Snyder}}, \bibinfo {author}
  {\bibfnamefont {G.}~\bibnamefont {Camy}}, \ and\ \bibinfo {author}
  {\bibfnamefont {M.}~\bibnamefont {Ducloy}},\ }\href {\doibase
  10.1103/PhysRevLett.44.1251} {\bibfield  {journal} {\bibinfo  {journal}
  {Phys. Rev. Lett.}\ }\textbf {\bibinfo {volume} {44}},\ \bibinfo {pages}
  {1251} (\bibinfo {year} {1980})}\BibitemShut {NoStop}%
\bibitem [{\citenamefont {Shirley}(1982)}]{Theory2}%
  \BibitemOpen
  \bibfield  {author} {\bibinfo {author} {\bibfnamefont {J.~H.}\ \bibnamefont
  {Shirley}},\ }\href {\doibase 10.1364/OL.7.000537} {\bibfield  {journal}
  {\bibinfo  {journal} {Opt. Lett.}\ }\textbf {\bibinfo {volume} {7}},\
  \bibinfo {pages} {537} (\bibinfo {year} {1982})}\BibitemShut {NoStop}%
\bibitem [{\citenamefont {Camy}\ \emph {et~al.}(1982)\citenamefont {Camy},
  \citenamefont {Bordé},\ and\ \citenamefont {Ducloy}}]{Theory3}%
  \BibitemOpen
  \bibfield  {author} {\bibinfo {author} {\bibfnamefont {G.}~\bibnamefont
  {Camy}}, \bibinfo {author} {\bibfnamefont {C.}~\bibnamefont {Bordé}}, \ and\
  \bibinfo {author} {\bibfnamefont {M.}~\bibnamefont {Ducloy}},\ }\href
  {\doibase https://doi.org/10.1016/0030-4018(82)90406-0} {\bibfield  {journal}
  {\bibinfo  {journal} {Optics Communications}\ }\textbf {\bibinfo {volume}
  {41}},\ \bibinfo {pages} {325} (\bibinfo {year} {1982})}\BibitemShut
  {NoStop}%
\bibitem [{\citenamefont {Schenzle}\ \emph {et~al.}(1982)\citenamefont
  {Schenzle}, \citenamefont {DeVoe},\ and\ \citenamefont {Brewer}}]{Theory4}%
  \BibitemOpen
  \bibfield  {author} {\bibinfo {author} {\bibfnamefont {A.}~\bibnamefont
  {Schenzle}}, \bibinfo {author} {\bibfnamefont {R.~G.}\ \bibnamefont {DeVoe}},
  \ and\ \bibinfo {author} {\bibfnamefont {R.~G.}\ \bibnamefont {Brewer}},\
  }\href {\doibase 10.1103/PhysRevA.25.2606} {\bibfield  {journal} {\bibinfo
  {journal} {Phys. Rev. A}\ }\textbf {\bibinfo {volume} {25}},\ \bibinfo
  {pages} {2606} (\bibinfo {year} {1982})}\BibitemShut {NoStop}%
\bibitem [{\citenamefont {Le~Boiteux}\ \emph {et~al.}(1984)\citenamefont
  {Le~Boiteux}, \citenamefont {Bloch}, \citenamefont {Ducloy}, \citenamefont
  {Pesnelle}, \citenamefont {Runge},\ and\ \citenamefont {Perdrix}}]{DOTMTS1}%
  \BibitemOpen
  \bibfield  {author} {\bibinfo {author} {\bibfnamefont {S.}~\bibnamefont
  {Le~Boiteux}}, \bibinfo {author} {\bibfnamefont {D.}~\bibnamefont {Bloch}},
  \bibinfo {author} {\bibfnamefont {M.}~\bibnamefont {Ducloy}}, \bibinfo
  {author} {\bibfnamefont {A.}~\bibnamefont {Pesnelle}}, \bibinfo {author}
  {\bibfnamefont {S.}~\bibnamefont {Runge}}, \ and\ \bibinfo {author}
  {\bibfnamefont {M.}~\bibnamefont {Perdrix}},\ }\href@noop {} {\bibfield
  {journal} {\bibinfo  {journal} {Optics communications}\ }\textbf {\bibinfo
  {volume} {52}},\ \bibinfo {pages} {274} (\bibinfo {year} {1984})}\BibitemShut
  {NoStop}%
\bibitem [{\citenamefont {Boiteux}\ \emph {et~al.}(1986)\citenamefont
  {Boiteux}, \citenamefont {Bloch},\ and\ \citenamefont {Ducloy}}]{DOTMTS2}%
  \BibitemOpen
  \bibfield  {author} {\bibinfo {author} {\bibfnamefont {S.}~\bibnamefont
  {Boiteux}}, \bibinfo {author} {\bibfnamefont {D.}~\bibnamefont {Bloch}}, \
  and\ \bibinfo {author} {\bibfnamefont {M.}~\bibnamefont {Ducloy}},\ }\href
  {\doibase 10.1051/jphys:0198600470103100} {\bibfield  {journal} {\bibinfo
  {journal} {Journal de Physique}\ }\textbf {\bibinfo {volume} {47}},\ \bibinfo
  {pages} {31} (\bibinfo {year} {1986})}\BibitemShut {NoStop}%
\bibitem [{\citenamefont {de~Escobar}\ \emph {et~al.}(2015)\citenamefont
  {de~Escobar}, \citenamefont {\'{A}lvarez}, \citenamefont {Coop},
  \citenamefont {Vanderbruggen}, \citenamefont {Kaczmarek},\ and\ \citenamefont
  {Mitchell}}]{DOTMTS3}%
  \BibitemOpen
  \bibfield  {author} {\bibinfo {author} {\bibfnamefont {Y.~N.~M.}\
  \bibnamefont {de~Escobar}}, \bibinfo {author} {\bibfnamefont {S.~P.}\
  \bibnamefont {\'{A}lvarez}}, \bibinfo {author} {\bibfnamefont
  {S.}~\bibnamefont {Coop}}, \bibinfo {author} {\bibfnamefont {T.}~\bibnamefont
  {Vanderbruggen}}, \bibinfo {author} {\bibfnamefont {K.~T.}\ \bibnamefont
  {Kaczmarek}}, \ and\ \bibinfo {author} {\bibfnamefont {M.~W.}\ \bibnamefont
  {Mitchell}},\ }\href {\doibase 10.1364/OL.40.004731} {\bibfield  {journal}
  {\bibinfo  {journal} {Opt. Lett.}\ }\textbf {\bibinfo {volume} {40}},\
  \bibinfo {pages} {4731} (\bibinfo {year} {2015})}\BibitemShut {NoStop}%
\bibitem [{\citenamefont {Lim}\ \emph {et~al.}(2022)\citenamefont {Lim},
  \citenamefont {McPoyle},\ and\ \citenamefont {Cervantes}}]{DOTMTS4}%
  \BibitemOpen
  \bibfield  {author} {\bibinfo {author} {\bibfnamefont {M.~J.}\ \bibnamefont
  {Lim}}, \bibinfo {author} {\bibfnamefont {S.}~\bibnamefont {McPoyle}}, \ and\
  \bibinfo {author} {\bibfnamefont {M.}~\bibnamefont {Cervantes}},\ }\href
  {\doibase https://doi.org/10.1016/j.optcom.2022.128651} {\bibfield  {journal}
  {\bibinfo  {journal} {Optics Communications}\ }\textbf {\bibinfo {volume}
  {522}},\ \bibinfo {pages} {128651} (\bibinfo {year} {2022})}\BibitemShut
  {NoStop}%
\bibitem [{\citenamefont {Zhang}\ \emph {et~al.}(2014)\citenamefont {Zhang},
  \citenamefont {Liu}, \citenamefont {Tao}, \citenamefont {Ling},\ and\
  \citenamefont {Chen}}]{velocity}%
  \BibitemOpen
  \bibfield  {author} {\bibinfo {author} {\bibfnamefont {L.-G.}\ \bibnamefont
  {Zhang}}, \bibinfo {author} {\bibfnamefont {Z.-Z.}\ \bibnamefont {Liu}},
  \bibinfo {author} {\bibfnamefont {Z.-M.}\ \bibnamefont {Tao}}, \bibinfo
  {author} {\bibfnamefont {L.}~\bibnamefont {Ling}}, \ and\ \bibinfo {author}
  {\bibfnamefont {J.-B.}\ \bibnamefont {Chen}},\ }\href {\doibase
  10.1088/0256-307X/31/8/083101} {\bibfield  {journal} {\bibinfo  {journal}
  {Chinese Physics Letters}\ }\textbf {\bibinfo {volume} {31}},\ \bibinfo
  {pages} {083101} (\bibinfo {year} {2014})}\BibitemShut {NoStop}%
\bibitem [{\citenamefont {Demtröder}(2015)}]{hole}%
  \BibitemOpen
  \bibfield  {author} {\bibinfo {author} {\bibfnamefont {W.}~\bibnamefont
  {Demtröder}},\ }\href {\doibase 10.1007/978-3-662-44641-6} {\emph {\bibinfo
  {title} {Laser spectroscopy 2}}}\ (\bibinfo {year} {2015})\ p.~\bibinfo
  {pages} {92}\BibitemShut {NoStop}%
\bibitem [{\citenamefont {Bloch}\ \emph {et~al.}(1982)\citenamefont {Bloch},
  \citenamefont {Raj}, \citenamefont {Peng},\ and\ \citenamefont
  {Ducloy}}]{bloch1982dispersive}%
  \BibitemOpen
  \bibfield  {author} {\bibinfo {author} {\bibfnamefont {D.}~\bibnamefont
  {Bloch}}, \bibinfo {author} {\bibfnamefont {R.}~\bibnamefont {Raj}}, \bibinfo
  {author} {\bibfnamefont {K.}~\bibnamefont {Peng}}, \ and\ \bibinfo {author}
  {\bibfnamefont {M.}~\bibnamefont {Ducloy}},\ }\href@noop {} {\bibfield
  {journal} {\bibinfo  {journal} {Physical Review Letters}\ }\textbf {\bibinfo
  {volume} {49}},\ \bibinfo {pages} {719} (\bibinfo {year} {1982})}\BibitemShut
  {NoStop}%
\end{thebibliography}%


%

\end{document}